\documentclass[12pt, preprint]{aastex}
\def\pppm{\rm P^3M}
\def\mpchi{\,h^{-1}{\rm {Mpc}}}
\def\kpchi{\,h^{-1}{\rm {kpc}}}
\def\kms{\,{\rm {km\, s^{-1}}}}
\def\msun{{\rm M_\odot}}
\def\msunhi{\,h^{-1}{\rm M_\odot}}

\begin{document}

\title{Semi-analytical model of galaxy formation with high-resolution N-body simulations}
\author{X. Kang$^{1,2}$, Y. P. Jing$^{1}$, H. J. Mo$^{3}$ and G. B\"orner$^{4}$}
\affil{$^1$
Shanghai Astronomical Observatory, the Partner Group
of MPI f\"ur Astrophysik, Nandan Road 80, Shanghai, China}
\affil{$^2$
Graduate School of the Chinese Academy of Sciences, 19A, Yuquan Road, Beijing, China}
\affil{$^3$
Department of Astronomy, University of Massachusetts,
Amherst MA 01003-9305, USA}
\affil{$^4$
Max-Planck-Institut f\"ur Astrophysik
Karl-Schwarzschild-Strasse 1, 85748 Garching, Germany}
\affil{e-mail: kangx@shao.ac.cn}

\begin{abstract}
We model the galaxy formation in a series of high-resolution N-body
simulations using the semi-analytical approach.  Unlike many earlier
investigations based on semi-analytical models, we make use of the
subhalos resolved in the $N$-body simulations to follow the mergers of
galaxies in dark halos, and we show that this is pivotal in modeling
correctly the galaxy luminosity function at the bright end and the
bimodal nature of galaxy color distribution. Mergers of galaxies based
on subhalos also result in many more bright red galaxies at high
$z$. The semi-analytical model we adopt is similar to those used in
earlier semi-analytical studies, except that we consider the effect of
a prolonged cooling in small halos and that we explicitly follow the
chemical enrichment in the interstellar medium. We use our model to
make predictions for the properties of the galaxy population at low
redshift and compare them with various current observations.  We find
that our model predictions can match the luminosity functions of
galaxies in various wavebands redder than the u-band. The shape of the
luminosity function at bright end is well reproduced if galaxy mergers
are modeled with the merger trees of subhalos and the steep faint-end
slope can be moderated if the gas cooling time in low-mass halos is
comparable to the age of the universe. The model with subhalos resolved
can reproduce the main features in the observed color
bimodal distribution, though it still predicts too many bright blue
galaxies. The same model can also match the color-magnitude relation
for elliptical galaxies in clusters, the metallicity-luminosity
relation and metallicity-rotation velocity relation of spiral
galaxies, and the gas fraction in present-day spiral galaxies. We also
identify areas where further improvements of the model are required.
\end{abstract}


\section{Introduction}
The recent observations of the Wilkinson Microwave Anisotropy Probes
(WMAP, Spergel et al. 2003), combined with many other observations on
large scale structures, favor a flat universe in which the dark
energy, $\Lambda$, dominates the expansion and evolution of the
Universe, and most of its non-relativistic matter is Cold Dark Matter
(CDM). The ordinary matter, i.e. the baryonic matter, accounts for
only a small fraction ($\sim 15\%$ of the non-relativistic matter).
With these favorable cosmological parameters plus some other
reasonable parameters for the Hubble constant and for the primordial
density fluctuations, the CDM model can match most of the current
observations, including the cosmic microwave background, the
intergalactic medium at high redshift, the abundance of rich clusters,
and the large-scale distributions of galaxies in the Two Degree Field
Galaxy Redshift Survey (2dFGRS, Colless et al. 2001), in the 2-Micron
All Sky Survey (2MASS, Skrutskie et al. 1997), and in the Sloan
Digital Sky Survey (SDSS, York et al. 2000) (e.g. Peacock et al. 2001;
Maller et al. 2003; Tegmark et al. 2004).  However, when we compare
theoretical models (such as the CDM model) with the galaxy
distribution revealed by large redshift surveys of galaxies, we have
to contend with the bias in the relation between the distribution of
luminous galaxies and the underlying dark matter. To understand this
relation is therefore one of the main challenges in modern cosmology.
With accurate multi-band photometries and medium resolution spectra
now available from large redshift surveys such as the 2dFGRS, the SDSS
and the DEEP2 Galaxy Redshift Survey (Coil et al. 2004), it is now possible to 
study in
detail a wide range of properties of the galaxy population, such as
spatial clustering on different scales, the dependence of clustering
on luminosity and color, the luminosity function, the magnitude-color
relation, and the environmental dependence of the galaxy
population. These properties can provide important clues to how the
galaxies have formed, and important constraints on the underlying
cosmological model, such as the primordial density fluctuation,
cosmological parameters, and the properties of the dark matter.

To fully understand the galaxy properties and to fully make use of
large surveys of galaxies to constrain cosmological models, it is
necessary to understand how galaxies form in the cosmic density
field. However, it is still a challenge to model galaxy formation
within the framework of CDM models, though significant progress has
been achieved in the last two decades. The parameter space for the
Big-Bang cosmology has now been narrowed greatly, and the
uncertainties caused by these parameters are relatively small.  Using
high-resolution $N$-body simulations and sophisticated analytical
models, the properties of the dark matter distribution are well
understood in the CDM scenario. In particular, a great deal have been
learned about the properties of the CDM halo population, which are
virialized clumps formed through gravitational instability in the
cosmic density field and in which galaxies are assumed to form. The
challenge for current galaxy formation is really to understand the
physical processes that govern galaxy formation and evolution in dark
matter halos, such as shock heating of gas, radiative cooling, star
formation, AGN activity and their feedback, and galaxy merging.

There are several ways to link galaxies to dark matter halos in a
given cosmological model. The most straightforward way is to simulate
galaxy formation in an expanding universe by numerically solving the
gravitational and hydrodynamical equations (e.g. Katz \& Gunn 1991;
Cen \& Ostriker 1993; Bryan et al. 1994; Navarro \& White 1994;
Couchman et al. 1995; Abel et al. 1997; Weinberg et al. 1998;
Yoshikawa et al. 2000; Springel et al. 2001a).  The hydro/N-body
simulation has the advantage of treating the gas dynamical processes
in a self-consistent manner. Other known important physical processes,
such as star formation and its feedback, are usually input into the
simulation ``by hand''. In order to resolve galaxies as well as to
include the effect of the large scale structures, these simulations
are required to cover a large dynamical range. With the recent rapid
developments of the algorithm, of the physical modeling, and of the
computer hardwares, the prospects of developing the hydro/N-body
simulations are very promising. But with the current technology, it is
still very time-consuming to run a realistic hydro/N-body simulation
of galaxy formation(e.g. Springel \& Hernquist, 2003), which makes it
difficult to study how the galaxy properties change with model
parameters.

Another powerful tool to link galaxies to dark matter halos is the
so-called halo occupation model (Yang et al. 2003; van den Bosch et
al. 2003; Berlind et al. 2003; Zheng et al. 2004; Vale \& Ostriker
2004; Kravtsov et al. 2004; for earlier works, see Jing, Mo, B\"orner
1998; Peacock \& Smith 2000; Seljak 2001; Berlind \& Weinberg 2002;
Bullock, Wechsler \& Somerville 2002; Cooray \& Sheth 2002). The halo
occupation model assumes a parameterized form for the conditional
luminosity function (CLF) that quantifies the luminosity distribution
of galaxies in a halo of mass $m$. In contrast to the hydro/N-body
simulations that aim to model galaxies from basic principles of
physics, the parameterized form of the halo model is motivated by
observations and the model parameters are determined by best-fitting
the observations such as the luminosity function of galaxies,
luminosity and color dependences of galaxy clustering, the
mass-to-light ratios of galaxy systems (Yang et al. 2003; van den
Bosch et al. 2003; Yan et al. 2003; van den Bosch et al. 2004). In the
future, more observational data can be incorporated to further refine
the halo occupation model, which will provide invaluable insight on
how galaxies form in CDM halos. However, the halo model itself does
not tell us about how the observed properties of galaxies have
developed through the cosmological evolution.
 
 A third method, which has been widely used and will be used here in
this paper, is to construct semi-analytical models (SAMs) of galaxy
formation (e.g., White \& Frenk 1991; Lacey \& Silk 1991; Kauffmann et
al. 1993; Cole et al. 1994; Somerville \& Primack 1999; Cole et
al. 2000).  The SAM approach lies in between the two methods described
above.  It incorporates physical processes that are well-known on the
basis of theoretical and observational studies, and parameterizes less
well understood physical processes by simple functional forms.  In
this way, a large number of physical processes can be implemented into
SAMs to a certain degree of accuracy, including hierarchical growth of
dark halos, shock heating of the intergalactic gas, radiative cooling
of the hot gas in halos, disk formation from cold gas, star formation
and its feedback, metal enrichment, starbursts, morphology
transformations associated with the mergers of galaxies, and dust
extinction. The parameters involved in the model are determined by
matching some well-established observations, e.g., the Tully-Fisher
relation, and the local luminosity functions of galaxies.  With star
formation explicitly included in the modeling and the physical
properties predicted for individual galaxies, SAM is powerful in
making model predictions that can be directly compared with
observations.  In the last two decades, a number of groups have used
the semi-analytical approach to interpret observations to constrain
theoretical models (e.g., White \& Rees 1978; White \& Frenk 1991;
Cole 1991; Lacey \& Silk 1991; Kauffmann et al. 1993; Cole et
al. 1994; Kauffmann et al. 1999a, hereafter KCDW; Somerville \&
Primack 1999; Cole et al. 2000; Nagashima et al. 2002; Menci et
al. 2002; Benson et al. 2003; Hatton et al. 2003, Helly et al. 2003).

One advantage of the semi-analytic approach is that it can be
combined with the merger trees of dark matter halos obtained directly
from $N$-body simulations to produce model galaxy catalogs (e.g. KCDW;
Springel et al. 2001b, hereafter SWTK; Hatten et al. 2003; Helly et
al. 2003; De Lucia et al. 2004).  Since such catalogs contain
information not only about the physical properties of individual
galaxies, but also about galaxy distributions in the phase space, they
are very useful in making comparisons between model predictions and
observations, and in generating mock catalogs to quantify
observational bias.  An example here is the GIF galaxy catalogs
produced by KCDW using their SAM, which have been used to study the
clustering properties of various types of galaxies at $z=0$ (KCDW) and
at high redshift (Kauffmann et al. 1999b), to produce mock catalogs
for the CfA2 redshift survey (Diaferio et al. 1999), and to interpret
the galaxy-galaxy lensing observation of the SDSS (Yang et al. 2003).

Despite the large effort, there are still outstanding problems in the
SAMs. The first problem is related to the fact that some important
physical processes, such as the feedback from star formation, are
still poorly understood, and so there is still a large freedom in the
implementations of such processes.  Secondly, even with the large
number of models that have been investigated, so far we still do not
have a single model that can match all the important observational
constraints.  For example, most of the SAMs studied so far have
difficulties in simultaneously matching the observed Tully-Fisher
relation and the luminosity function of galaxies; it is also difficult
for the current SAMs to simultaneously account for the observed
luminosity density at the present time and the star formation rate at
$z\ga 3$ as inferred from the sub-mm sources (Chapman et al. 2002;
Cowie et al.  2002; Smail et al. 2002; Borys et al. 2003; Baugh et
al. 2004).  Other problems, such as the faint-end slope of the
luminosity function of galaxies, and the sharp break of the galaxy
luminosity function at the bright end, still lack satisfactory
solutions (e.g. Benson et al. 2003). Clearly, further investigations
are required within the framework of the SAM.

One significant recent development in the studies of structure
 formation is the finding that CDM halos are not smooth objects, but
 contain many subhalos instead (e.g. Jing et al. 1995; Ghigna et
 al. 1998; Moore et al. 1999; Klypin et al. 1999; SWTK). Since
 galaxies may have formed in the centers of subhalos at high
 redshifts, the properties of the subhalo population may provide
 important clues about the formation and evolution of galaxies in dark
 matter halos.  Furthermore, following the motion of subhalos can
 provide more precise information about the positions and velocities
 of the galaxies hosted by subhalos, which is important for the
 comparison between theoretical predictions and the observational
 results derived from redshift surveys. The results about the subhalo
 population have yet to be fully incorporated into the SAM. One step
 in this direction has been taken by SWTK and De Lucia et al. (2004)
 who used their high-resolution halo simulations together with SAM to
 study the formation and evolution of galaxies. The halos in their
 simulations have masses typically of a cluster of galaxies, and are
 simulated with a multi-mass particle technique, so that subhalos are
 well resolved. They identified subhalos in the simulation and
 followed explicitly the formation and evolution of galaxies in the
 subhalos with the SAM. One of the most interesting results they
 obtained is that the subhalo-based SAM can significantly improve the
 agreement between the model prediction and the observation for the
 cluster galaxy luminosity function. The main reason is that the
 merger time scale adopted in previous SAMs underestimated the merging
 time for bright galaxies (with luminosity $\sim L_\star$, the
 characteristic luminosity of the Schechter form). This
 underestimation leads to a reduced survival time for bright galaxies
 (thus the number of bright galaxies), resulting in central galaxies
 that are too bright, and in a cluster luminosity function that is
 much less curved around $L_\star$ than the observed one.

 In this paper, we consider galaxy formation in a typical cosmological 
volume, combining SAM with a set of high-resolution N-body simulations
that can resolve subhalos in massive halos. Our aim is two-fold.
Firstly, we want to examine whether resolving subhalos in massive 
halos can also improve model predictions for the luminosity function
of the overall galaxy population, or one has to resolve low-mass halos
in order to model the overall luminosity function correctly.  
Secondly, we want to use a variety of current observational results,
including the multi-waveband luminosity functions and 
color distribution, to constrain the current SAMs and to identify 
issues for which further improvements of the model has to be made. 
The arrangement of our paper is as follows.  In Section 2, we present the
simulations and briefly introduce our algorithm to find subhalos. In
Section 3, we describe the main physical processes and how to implement
them in our SAM model. In Section 4, we test the model results by comparing
with a handful of well-established observations of galaxies. In Sec 5, we
summarize and discuss our main results.

\section{$N$-body simulations and subhalo merging trees}

\subsection{The cosmological model and $N$-body simulations}

\begin{figure}
\plotone{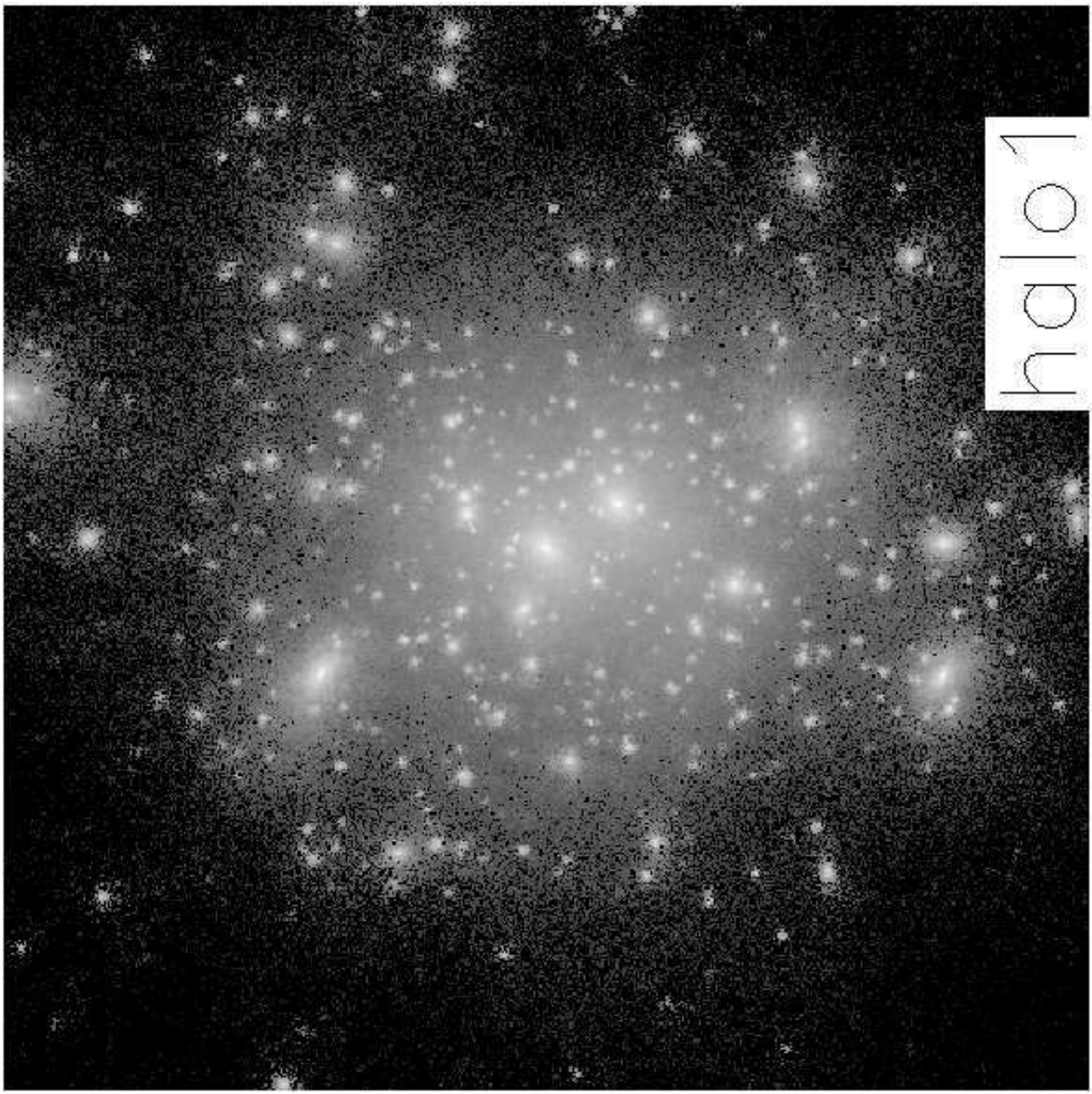} 
\caption{Dark matter distribution within the most massive halo in the 
LCDM100 simulation. The physical size of the figure is 2 times  the halo 
virial radius. }
\label{fig:halo}
\end{figure}

The main simulation used in this paper is a $\pppm$ cosmological
simulation of $512^3$ particles in a box of $100\mpchi$. The
underlying cosmological model is the standard concordance model with
the density parameter $\Omega_{\rm m,0}=0.3$ , the cosmological
constant $\Omega_{\Lambda,0}=0.7$ and the Hubble constant $h=0.7$ in
units of $100\kms {\rm Mpc}^{-1}$. The initial density field is
assumed to be Gaussian with a Harrison-Zel'dovich primordial power
spectrum and with an amplitude specified by $\sigma_8=0.9$, where
$\sigma_8$ is the r.m.s. fluctuation of the linearly evolved density
field in a sphere of radius $8\mpchi$.  This simulation, which started
at redshift $z_i=72$, is evolved by 5000 time steps to the present
$z=0$ with our vectorized parallel $\pppm$ code (Jing \& Suto 2002) at
the VPP5000 Fujitsu supercomputer of the National Astronomical
Observatory of Japan. The force softening length $\eta_f$ (S2 type,
Hockney \& Eastwood 1981) is $10\kpchi$ comoving, and the particle
mass $m_p=6.2\times 10^{8}\msunhi$. Because these simulation
parameters are very similar to those adopted in many high-resolution
re-simulations of individual cluster halos (Moore et al. 1999, Jing \&
Suto 2000, Fukushige \& Makino 2001, 2003, Power et al. 2003, Diemand
et al 2004), we have achieved a resolution that can resolve subhalos
within massive halos. As we will see below, this is crucial for
modeling correctly the mergers of galaxies in dark halos.  The
over-merging problem, which exists in many previous studies of
structures of dark matter halos and semi-analytic models of galaxy
formation based on cosmological $N$-body simulations, is greatly
alleviated in our model. As an illustration, we show in Figure
\ref{fig:halo} the density distribution of dark matter particles
within the largest dark halo in the simulation. As one can see, many
subhalos remain intact during the evolution. We have estimated the
mass function of the subhalo population (defined in
\S\ref{sec:subhalo}), and found a good agreement with the subhalo mass
functions obtained in previous halo re-simulations (e.g. SWTK) down to
a mass of about $6.2\times 10^{10} \msunhi$ (100 particles).  This
simulation was used to derive the triaxial model for dark matter halos
in Jing \& Suto (2002), to which we refer the reader for some
complementary information about the simulation. We will call this
simulation L100 in the following.

Although a box of $100\mpchi$ usually can be regarded as a 
typical volume in the universe, the most massive clusters 
with masses larger than $\sim 10^{15}\msunhi$ may still be
under-represented,  because of the wavelength cutoff beyond 
the simulation box. Therefore, we have carried out 
simulations separately for 20 such massive halos with the 
nested-grid $\pppm$ code of Jing \& Suto (2000). These massive
halos are picked up from a cosmological simulation of $256^3$
particles in a box of $300\mpchi$, and they are the most 
massive halos in the simulation. We do not impose
any other criterion on the selection of these halos, 
and so the halo sample should be unbiased. In the 
re-simulations, each of the massive halos is represented
by about $2\times 10^6$ particles within the virial
radius, and so the particle mass is $(0.5\sim 1) \times
10^9\msunhi$,  comparable to the particle mass in L100. The force
softening is also about $10\kpchi$, but the simulations have been
evolved by 20000 time steps (this choice of time steps
was later found to be too conservative). This sample will be 
used to populate massive halos with galaxies in a set
of simulations of $300\mpchi$ box in a subsequent paper. Here we only use
the first halo $CL1$ to compare our results for cluster galaxies 
with observations and with those obtained by Springel et al. (2001). 
In addition to those simulations, we have also a cosmological simulation 
of $256^3$ particles in a box of $25\mpchi$, which has $\eta_f=5\kpchi$ 
and is evolved with 5000 time steps. This simulation
(referred to as L25 in the following)  
has a better mass resolution (8 times smaller in particle mass) than 
the L100 simulation. We will use this simulation to examine the mass 
resolution effect on the properties of galaxies at the faint luminosity 
end. All the three sets of simulations have the same model parameters. 
The L100 and CL simulations have 60 outputs from $z=15$ equally spaced in 
$\log_{10} (1+z)$, and the L25 has 165 outputs. These outputs
are used to construct the merging trees of dark halos. 

\subsection{Merger trees of subhalos}
\label{sec:subhalo}

The dark matter halos are identified in the simulations using the
Friends-of-Friends method (FOF) with a linking length equal to 0.2
of the mean particle separation. The technique rarely breaks a
physically bound halo into pieces; rather, it may combine two nearby
distinct halos into one halo in some cases, if there is a thin bridge
of particles between the two halos. Furthermore, in our high
resolution simulations, galactic halos falling into a more
massive halo (say, a cluster halo) can survive the tidal disruption
of the massive halo for a considerable period of time. 
As emphasized in SWTK, it is important to follow the trajectories 
of these subhalos, not only because they can provide us with the 
position and velocity of the galaxies they host, but also 
because they allow us to accurately follow the merging
histories of galaxies (as discussed in \S1). We identify the subhalos within
the FOF halos with the SUBFIND routine of SWTK, which 
kindly has been made available to us by Volker Springel.

The SUBFIND routine finds self-bound halos and subhalos in a single output of
a simulation.  Starting from a FOF halo, the SUBFIND first locates
local overdense regions as subhalo candidates. For each particle
in the input FOF halo, the local density is computed using the common
SPH technique. The smoothing length is taken to be the distance to the
$N_{\rm dens}$-th nearest neighbor particle, and the density is
smoothed by a kernel interpolation of the $N_{\rm dens}$ particles.
The particles are sorted in order of decreasing density, and are
assigned to subhalo candidates sequentially. When the $i$-th particle
with density $\rho_{\rm i}$ is considered, a set of particles $A_{i}$
is defined as the set of its $N_{\rm ngb}$ 
nearest neighbors. In the set of
$A_{i}$ particles, we select a subset of particles with density higher
than $\rho_{i}$ and among those a set $B_{i}$ is selected as the set
of the
two closest particles. Particle $i$ is assigned to a subhalo candidate
according to the set of particles $B_{i}$. If there is no particle in
$B_{i}$, particle $i$ is at the local maximum of the smoothed density
field, and a new subhalo candidate is created around particle $i$. If
there is only one particle $j$ in $B_{i}$, the particle $i$ is likely
within the subhalo that contains particle $j$, and thus is assigned to
that subhalo candidate. If there are two particles $j$ and $k$ in
$B_{i}$, there are two possibilities. Either the two particles $j$ and
$k$ are in two different subhalos. In this case, the particle $i$ is
said to be a {\it saddle} point connecting these subhalos, and a new
subhalo candidate is formed by joining these two subhalos with
particle $i$. Or, as the second possibility, the two particles $j$ and
$k$ are in the same subhalo. In this case, particle $i$ is simply
added to this subhalo candidate. In this way, a set of subhalo
candidates are created for the FOF group. Then the particles in each
subhalo candidate are checked whether they are bound in the
center-of-mass reference frame, and the particles that have positive
total energy are removed iteratively until all the particles are
bound. The bound subhalo candidates thus constructed establish a
hierarchy of small subhalos in larger subhalos. 
The subhalos that have only one saddle point are at the highest 
level of the hierarchy, and those with 
more saddle points are at a lower level of hierarchy. 
In this case, one particle may be assigned more than
once to subhalos at different levels of the hierarchy. 
SUBFIND keeps the identity of the particles only at the highest 
possible level of the hierarchy, thus producing a unique subhalo 
catalog. The algorithm can easily deal with the subhalo-within-subhalo 
problem, as well as the problem of the FOF algorithm that two halos 
with a thin bridge of particles are identified as one halo. 
With these subhalo catalogs at different outputs, we construct 
the merging trees for subhalos.

\begin{figure}
\plotone{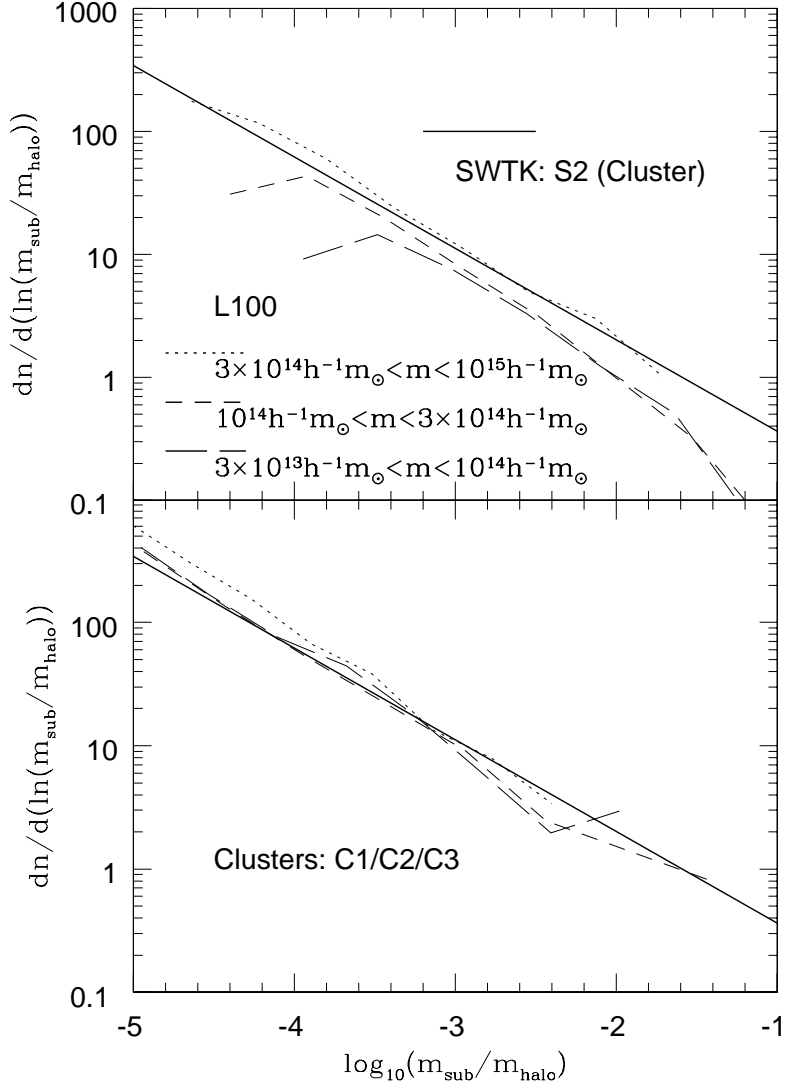} 
\caption{The mass function of subhalos as a
function of the mass ratio of the subhalo to the host halo. The
upper panel shows the mean result measured from the L100 simulation
for three mass ranges of host halos (as indicated in the panel). 
The lower panel shows the results for the first three clusters 
in the sample of individual halo re-simulations. 
All the results are shown down to a subhalo mass
corresponding to 16 simulation particles. For comparison, the mass
function of cluster halos obtained by SWTK from their halo
re-simulations is plotted as the solid line.  }
\label{submf}
\end{figure}

In Figure \ref{submf}, we show the mass functions of the subhalos in
the L100 simulation (upper panel) and in the cluster halo
re-simulations (low panel). For the L100 simulation, the results are
the average values for the host halos in the mass ranges indicated in
the figure. For the cluster halo re-simulations, the mass functions are
for individual halos. These mass functions are plotted down 
to a mass corresponding to 16 simulation particles. The mass functions 
for the most massive halos in the L100 simulation and for the individual 
cluster re-simulations agree well with each other
over the whole mass range. For comparison we show the prediction of SWTK by
the thick solid line. It was found that the mass function have a little steeper
slope than that of SWTK and close to the reported relation 
$n\propto m_{\rm sub}^{-2}$ by Diemand et al (2004).
The subhalo mass function for halos with masses 
$m_{\rm halo}=5\times 10^{13}\msunhi$ has a similar slope to that of
cluster halos, but the amplitude is a factor of about 2 lower. This
decrease in amplitude with host halo mass is quantitatively in
agreement with the recent results obtained by Gao et al. (2004)
based on a cosmological simulation that has slightly lower resolution
than our L100. The figure also shows that the subhalo mass function
breaks away from the power law, $n\propto m_{\rm
sub}^{-2}$, for subhalos containing less than about 20 simulation 
particles in the L100 simulation, indicating that the mass function 
may be affected by the resolution effect at such a mass scale. 
These results indicate that our L100 simulation can be used 
to probe subhalos that contain more than 50 simulation 
particles (i.e. with masses $\ga 3\times 10^{10}\msunhi$). 
For a subhalo with less than 50 particles, we use the L100 simulation 
to follow its merger history until it reaches the resolution limit, i.e. 10 
particles. We do not use the Monte Carlo method to follow the merger 
histories of low mass halos, but instead we will use
 the L25 simulation (with a higher mass resolution) to check the resolution effect. 
As we will show in Section 4.1, the resolution limit has little effect on 
the predicted luminosity function for $M_{g} - 5 \log h < -17.5$, but it under-predicts
the number count of faint galaxies ($-17.5 <M_{g} - 5\log h < -15.5$) by $\sim 30\%$.

\section{The galaxy population}

We implement the SAM in our $N$-body simulations following SWTK, 
but our model also includes the chemical evolutions of galaxies 
and the inter-galactic medium, as well as dust extinction effects on 
observational quantities, such as luminosity and color. 
Our model is therefore closer to the model adopted in a more recent 
publication from the same group (De Lucia et al. 2004), 
although there are differences in the choice of model parameters. 
For example,  De Lucia et al. have chosen a high yield of metal 
production in order to match the observed high metallicity of the 
intracluster medium (ICM). Since the high metallicity leads to rapid
gas cooling, they had to use a rather high feedback efficiency 
to suppress the formation of luminous galaxies. 
In our model, we adopt a metal yield and a feedback efficiency 
that are more moderate. For completeness, we briefly describe our SAM in 
this section. Since our main treatments about galaxy formation are similar to 
SWTK, we will compare in some detail our prescription and results 
with those in SWTK and related investigations (e.g. KCDW, De Lucia 2004). 

\subsection{Galaxy population in the subhalo scheme}

 In the subhalo scheme, there are three distinct populations of
galaxies that are associated with FOF halos.  The largest subhalo in
a FOF halo is called the {\it main} halo, and the main halo always
hosts a galaxy at its center, which we call the {\it central} galaxy. The
central galaxy is assumed to have the same position and velocity 
as the most bound particle of the main halo. Any other subhalo in 
the FOF halo also hosts a galaxy at its center. These galaxies, 
called {\it halo} galaxies, were central galaxies before their host 
halos fell into a larger halo. They are assigned the positions and 
velocities of the most bound particles of their host subhalos. 
The halo galaxies have a physical status very similar to that of the central
galaxies, except that the {\it central} galaxies are fed by 
gas cooling flows, while the {\it halo} galaxies are not. 
When two or more halos and/or subhalos merge, the smaller 
halo (subhalo) may be destroyed, thus losing its identity as a 
subhalo in the SUBFIND algorithm. A subhalo may
also be tidally disrupted after passing through dense regions in the
main halo. A galaxy that was a central or a halo galaxy 
at an earlier epoch, but whose host (sub)halos disappeared from 
the SUBFIND catalog at some later time, is attached to the 
most bound particle of its host (sub)halo at the time 
just before the (sub)halo is disrupted. Such galaxies are called 
{\it satellite} galaxies.

The transformation of galaxies from one population to another 
is governed by the subhalo merging trees constructed from the 
SUBFIND catalogs at different outputs of the simulation.  
Consider two successive outputs at redshifts $z_a$ and
$z_b$ ($z_a<z_b$). The subhalo $S_{b}$ at redshift $z_{b}$ is said to
be a progenitor of a subhalo $S_{a}$ at a later time $z_a$ if more
than half of the $N_{\rm link}$ most-bound particles of $S_{b}$ are
contained in $S_{a}$. As in SWTK, we take $N_{\rm link}$ to be 10. After
excluding those volatile subhalos at $z_b$ that are not a progenitor
to any subhalo or FOF group at the later time $z_a$, each
subhalo at $z_b$ must be a progenitor to one and only one subhalo at
$z_a$. A subhalo at $z_a$ may have more than one progenitor at $z_b$,
and the most massive of them is referred to as the main progenitor. In the
case that a subhalo at $z_a$ has no progenitor, the subhalo is usually
the main halo of a FOF group newly formed between $z_a$ and $z_b$, and
a {\it central} galaxy is created for the subhalo. For a subhalo at $z_a$
that has at least one progenitor, the {\it central} ({\it halo}) galaxy of its
main progenitor at $z_b$ is updated to be its {\it central} ({\it halo}) galaxy, and
all the other galaxies of the progenitor subhalos (including the
satellites of the main progenitor) become the {\it satellite} galaxies of
the subhalo at $z_a$. The satellite galaxies are assumed to 
merge later with the central (halo) galaxy of the subhalo-- 
after a dynamical friction timescale (see \S\ref{sec:merger} 
for details).

The evolution of galaxies is assumed to be driven by 
the physical processes to be described in detail in 
Section 3.2. We solve the differential equations governing 
galaxy evolution using a time step smaller than the time interval 
between two successive simulation outputs. Typically, 20 steps are 
used for each interval between two successive outputs.

In passing, we mention that the virial mass $m_{\rm vir}$ of the 
main halo is defined as the mass enclosed by the virial 
radius $R_{\rm vir}$, within which
the mean density is $\Delta_{c}(z)$ times the critical density of the
universe at the redshift $z$ in consideration. 
$\Delta_{c}(z)$ can be calculated analytically from the 
spherical collapse model (e.g. Kitayama \& Suto 1996). 
Here we use the fitting formula provided by  Bryan \& Norman (1998),
\begin{equation}
\Delta_{c}(z)=18\pi^{2}+82x-39x^{2}\,,
\end{equation}
where $x=\Omega(z)-1$, and $\Omega(z)$ is
the density parameter at redshift $z$. For a subhalo, the virial mass
is simply the total mass of all the particles in the subhalo, and the
dynamical time $t_{\rm dyn}$ and virial velocity $V_{\rm vir}$
are kept at the values the subhalo had just before it merged into a 
bigger halo.

\subsection{Physical Processes}

 In addition to the formation of the halo/subhalo population,  
galaxy formation and evolution also involve many other 
physical processes. In this paper, we attempt to take into account 
all processes that are known to be important for the formation and 
evolution of galaxies. These include: (1) radiative cooling of the hot 
gas within the main halo; (2) star formation in the cold gas, and energy 
feedback into the cold gas by supernova explosions; (3) chemical evolution 
in the cold and the hot gas, and in stars; (4) mergers of galaxies, star 
bursts, and the transformation of galaxy morphologies; 
(5) spectro-photometric evolution of individual galaxies; 
and (6) the effect of dust extinction on observational quantities.

\subsubsection{Cooling of hot gas}
\label{ssec:cooling}

As in White \& Frenk (1991), the hot gas within a halo is assumed to
be distributed in an isothermal sphere $\rho_{\rm g}(r)$ with a virial
temperature $T=35.9(V_{\rm vir}/\kms)^2{\rm K}$. The local cooling
time at a given radius $r$ is
\begin{equation}
t_{\rm cool}(r)=\frac {3} {2} \frac{k T \rho_{\rm g}(r)} {\overline{\mu}
m_{\rm p} n_{\rm e}^2(r)\, \Lambda(T,Z)}\,,
\end{equation}
where $\overline{\mu}m_{\rm p}$ is the mean particle mass, $n_{\rm
e}(r)$ the number density of electron at $r$, and $\Lambda(T,Z)$ 
the cooling function, which depends both on the temperature $T$ and 
the metallicity $Z$ of the hot gas. We use the cooling 
functions tabulated in Sutherland \& Dopita (1993). The chemical 
evolution of the hot gas is described in \S\ref{sec:chemical}.

As in KCDW, we define the cooling radius $r_{\rm cool}(z)$ as the
radius at which the cooling time $t_{\rm cool}$ is equal to the age 
of the universe $t_{\rm age}(z)$ at redshift $z$. 
If the cooling radius is smaller than the virial radius, 
the mass of gas cooled per unit time, i.e. the mass cooling rate, 
in the halo is approximated by  
\begin{equation}
\dot{m}_{\rm cool} =4\pi  \rho_{\rm g}(r_{\rm cool}) 
r^2_{\rm cool}\dot{r}_{\rm cool}\,,
\label{eq:rcoollh}
\end{equation}
where a dot denotes the derivative with respect to time $t$. 
For the isothermal sphere distribution considered here, the cooling 
rate can be written as 
\begin{equation}
\dot {m}_{\rm cool} = \frac {m_{\rm hot}} {R_{\rm vir}} \frac {r_{\rm cool}}
{2t_{\rm cool}}\,,
\label{eq:rcoolsh}
\end{equation}
where $m_{\rm hot}$ is the total hot gas in the halo. 
Note that this relation is derived under the assumption  
that the isothermal sphere is static. It implies that 
all the gas within the cooling radius can cool in a time 
interval that is twice the cooling time. In what follows
we will assume this relation to hold during the evolution 
of a relatively massive halo, with $m_{\rm hot}$ being the 
total amount of hot gas at the time in consideration.
For small halos or halos at high redshift, 
the cooling time can be smaller than the age of the universe
even at the virial radius. In such halos, all the gas within the
virial radius can cool, and so the mass cooling rate is limited by
the rate at which gas is accreted into the system, rather than
by radiative cooling. In this case, we write the mass 
`cooling' rate as
\begin{equation}\label{eq:rcoolsh1}
\dot {m}_{\rm cool} = \frac {m_{\rm hot}} {t_{\rm cool}} \,,
\label{eq:rcoollhalo}
\end{equation}
where $t_{\rm cool}$ is to be specified below. Note that 
for halos where $r_{\rm cool}= R_{\rm vir}$, there 
is a jump of a factor of 2 between equation (\ref{eq:rcoolsh})
and equation (\ref{eq:rcoolsh1}), which is entirely due to 
the simple prescription we adopt. In reality, the transition
must be smooth, which can be made by joining  
equations (\ref{eq:rcoolsh}) and (\ref{eq:rcoolsh1}) with a 
smooth function. This does not lead to any noticeable 
effects, and so we will ignore such details.

 We assume a universal baryon fraction $f_{b}$ for every halo,
and the hot gas available for cooling is:
\begin{equation}
m_{\rm hot} =
f_{b}m_{\rm vir}-
\sum_{i}[m_{\star}^{(i)}+m_{\rm cold}^{(i)}+m_{\rm eject}^{(i)}]
\label{eq:mhot}
\end{equation}
where $f_{b} = \Omega_{b}/\Omega_{0}$, and $m_{\rm \star}^{(i)}$,
$m_{\rm cold}^{(i)}$, and $m_{\rm eject}^{(i)}$ are, respectively, the
masses in stars, in cold gas, and ejected from stars.  The sum is over
all the galaxies in the halo in consideration.  Because the ejected
mass is assumed to be mixed into the hot gas in the current work, the
term $ m_{\rm eject}^{(i)}$ is set to be 0 in the above equation.

 SWTK defined the cooling radius by equating $t_{\rm cool}(r)$ to the
dynamical time of the halo, i.e., $t_{\rm cool} = t_{\rm dyn}= R_{\rm
vir}/V_{\rm vir}$. On the other hand, KCDM used $t_{\rm cool}=t_{\rm
age}$, i.e.  equation (\ref{eq:rcoolsh}) for large halos where $r_{\rm
cool}<R_{\rm vir}$, while $t_{\rm cool}=t_{\rm dyn}$ for smaller halos
(eq.\ref{eq:rcoolsh1}). Note that the cooling rate given by $t_{\rm
cool}=t_{\rm dyn}$ is a factor of 2 to 4 higher than that given by
$t_{\rm cool}=t_{\rm age}$, and so the two assumptions in general give
quite different results for the galaxy population.

 Unfortunately, there is no good physical reason to prefer
either of the two assumptions. Although radiative cooling 
is a well-understood process, the gas cooling rate in a halo
depends sensitively on the assumed density and temperature profiles 
of the gas. The assumption usually made in SAMs
is that the gas is heated by gravitational collapse to a temperature
close to the virial temperature of the dark halo and that 
the gas distribution in a dark halo is close to that 
of an isothermal gas in hydrostatic equilibrium in the halo potential. 
However the heating and cooling of gas in a dark halo 
are more complicated than this assumption implies.  
On the one hand, detailed analytical models and hydrodynamic simulations 
show that the halo gas in low-mass halos may never be heated to 
the virial temperature of the halo (e.g. Birnboim \& Dekel 2003; 
Keres et al. 2004), and so the amount of cold gas may be underestimated 
by the above formulae. On the other hand, if gas in protogalactic halos 
is preheated to some finite entropy, the gas distribution 
around a small dark matter halo may be much more extended, 
and the rate of gas cooling in such systems
can be significantly reduced (Mo \& Mao 2002; Oh \& Benson 2003). 
It has been shown by some SAMs that the suppression of gas cooling 
in small halos by photoionization heating may have significant
impact on the faint-end slope of the galaxy luminosity function
(e.g. Somerville 2002; Benson et al. 2002; Benson et al. 2003), but such models depend 
on the details of the implementation of the process.  
In this paper, instead of introducing such complicated physical models, 
we consider a simple model, in which $t_{\rm cool}$ is assumed to be 
equal to $t_{\rm age}(z)$. This assumption is consistent with the 
results of Brinchmann et al. (2004) who have shown that
the local star-forming population of galaxies have roughly 
constant star formation rates over a time scale comparable 
to the present age of the Universe. As we will see below, this 
assumption leads to sufficient suppression of star formation 
in low-mass halos. 

 It has been pointed out by Kauffmann et al. (1993) that gas cooling
(or star formation) must be strongly suppressed in the most massive
halos so as not to produce too many super-luminous galaxies (see also
Benson et al. 2003). As we will see below, the assumption about this
suppression has significant impact on the prediction of the number of
bright blue galaxies.  As in Kauffmann et al. (1993), we consider a
case in which gas cooling is simply switched off for halos with
circular velocity larger than $390\kms$. Since there is no solid
physical motivation for this assumption, we will discuss another
simple model for the shut-off of cooling in subsection 4.2.

\subsubsection{Star formation}

We model the star formation rate $\psi$ as :
\begin{equation}
\psi = \alpha m_{\rm cold}/t_{\rm d}\,.
\label{eq:sfr1}
\end{equation}
In the above equation, $\alpha$ is the star formation efficiency, and
$t_{\rm d}$ is the dynamical time of the galaxy (not the halo).
As in SWTK we use $t_{\rm d}=0.1\times R_{\rm vir}/V_{\rm vir}$. 
Note that other choices, such as a constant $t_{\rm d}$,
have also been used in the literature (Somerville \& Primack 1999).
For a satellite galaxy, we keep the dynamical time 
$t_{\rm d}$ to be equal to the value the galaxy had when it was
last a central or halo galaxy. The star formation efficiency $\alpha$ is
assumed to depend on the circular velocity of the galaxy (Cole
et al. 1994; 2000). Following De Lucia et al. (2004), we take
\begin{equation}
\alpha = \alpha_{\rm 0} \left(\frac {V_{\rm vir}} 
{220 \kms}\right)^{ n}\,,
\label{eq:sfr2}
\end{equation}
where $\alpha_{0}$ and $n$ are treated as free parameters. Note 
that $\alpha$ cannot be larger than $1$. We take 
$\alpha=1$ if the above equation gives a value larger than 1.
Note that $n=0$ gives a constant star formation efficiency,
which is in fact consistent with the observations of
high mass galaxies (e.g. Kennicutt 1998). On the other hand,
there is evidence that low-mass galaxies convert cold gas 
into stars less efficiently than high-mass galaxies 
(e.g. Boissier et al. 2001, Kauffmann et al. 2003). In this case, 
$n$ should be positive. As shown by De Lucia et al. (2004), 
the assumed value of $n$ has significant impact on the predicted
cold gas fraction and we note that $n$ also plays an important role 
in determining the metallicity and color of low-mass galaxies. 
In the present paper, we follow De Lucia et al. (2004) by 
setting $n=2.2$.

\subsubsection{Feedback}

For a given stellar initial mass function (IMF), the energy released
by supernovae associated with massive stars is $\eta_{\rm SN}E_{\rm
SN}$ for a unit solar mass of newly formed stars, where $E_{\rm SN}$
is $\sim 10^{51} {\rm erg}$, the typical kinetic energy of the ejecta from
each supernova, and $\eta_{\rm SN}$ is an factor that depends on the 
form of the IMF. For the Scalo IMF (Scalo 1986), $\eta_{\rm SN}$ is
about $5.0 \times 10^{-3}$, and for the Salpeter (1955) IMF it 
is about $6.3 \times 10^{-3}$ (Note that both cases are assumed  
to apply in the mass range $0.1M_{\odot} < m < 125M_{\odot}$). 
The released energy is supposed 
to heat the cold gas and expel it from the cold gas disk. 
With an amount of $\Delta m_{\star}$ of newly formed stars, 
the amount of cold gas that can be heated can be expressed as,
\begin{equation}
\Delta m_{\rm eject} = \frac {4} {3}\epsilon \frac {\eta_{\rm
SN}E_{\rm SN}} {V_{\rm vir}^{2}} \Delta m_{\star}\,,
\end{equation}
where $\epsilon$ describes the feedback efficiency. 
We assume that the cold gas is reheated to the virial
temperature of the subhalo in which the galaxy resides.  
There are at least two possibilities for the fates of the reheated gas. 
The first is that the reheated gas always stays in the halo and 
is added to the hot gas for later cooling. The SAM model based on this 
treatment of the reheated gas is usually called the {\em retention} model. 
Another possibility is that the reheated gas is ejected from the 
host halo.  This ejected gas may be re-accreted by the halo 
after some time, say, one dynamical time of the halo. 
The SAM based on the this assumption is called the {\em ejection} model. 
It has been shown by SWTK that both the {\em retention} and 
the {\em ejection} models can be made to match most observations 
by adjusting model parameters, such as the star formation efficiency
$\alpha$ and the feedback parameter $\epsilon$. In this paper,
we only consider the {\em retention} model.
Note that in both retention and ejection cases, there is the issue 
about the state of the heated gas when it is re-accreted.
Usually, the re-heated gas is assumed to have completely forgot 
its thermal history when it is re-accreted. In reality, the heated 
gas may retain part of the entropy it gained and behave differently 
from the un-heated gas. Unfortunately, this process has not yet been 
implemented reliably in the current SAMs. 
  
\subsubsection{Chemical evolution}
\label{sec:chemical}

In the previous subsections, we have described the physical processes
that govern the mass and metal exchange among the three
main components,  stars, cold gas, and hot gas, within a dark 
matter halo. Here we write down the equations describing these 
processes, so that we can follow the evolution of each individual
component. We assume that a fraction $R$ of the mass in newly 
formed stars is recycled into the cold gas through stellar winds. 
For the Scalo IMF, $R\approx 0.4$, and for the Salpeter IMF, 
$R\approx 0.35$, according to the standard stellar population 
synthesis model (Cole et al. 2000). Thus, the change of the stellar 
mass in each galaxy is given by
\begin{equation}
\dot{m}_{\star} = (1 - {\rm R}) \psi\,.
\label{eq:mstars}
\end{equation}
The cold gas in a galaxy is increased by radiative cooling
of the hot gas, and is reduced by the formation of new stars 
and by supernova feedback. Thus, the change in the cold gas 
in each galaxy is given by
\begin{equation}
\dot{m}_{\rm cold} = \dot{m}_{\rm cool} - (1 - {\rm R})
\psi -\dot{m}_{\rm eject}\,.
\end{equation}
In our model, $\dot{m}_{\rm cool}=0$ for halo and satellite 
galaxies. In the retention model adopted in this paper, all the cold gas 
reheated by the galaxies is retained in the main halo, thus the 
change in the hot gas in the main halo is
\begin{equation}
\dot{m}_{\rm hot} = -\dot{m}_{\rm cool} + \sum_{i} \dot{m}_{\rm
eject}^{(i)}\,,
\end{equation}
where the summation is over all galaxies in the FOF halo.
 Now consider the metal contents in the three components. 
We assume a yield $p$ of metal for the conversion of a
unit mass of cold gas into stars. All the metals produced
are assumed to be returned to the cold gas instantaneously. 
Thus, the mass changes in heavy elements in
the three components are respectively given by
\begin{equation}
\dot{m}_{\star}^{Z} = (1-R)Z_{\rm cold}\psi\,,
\end{equation}
\begin{equation}
\dot{m}_{\rm cold}^{Z} = \dot{m}_{\rm cool}Z_{\rm hot} + p  \psi -
(1  - R)Z_{\rm cold} \psi - \dot{m}_{eject} Z_{\rm cold}\,,
\label{eq:coldmetal}
\end{equation}
and
\begin{equation}
\dot{m}_{\rm hot}^{Z} = -\dot{m}_{\rm cool} Z_{\rm hot}+\sum_{i}
\dot{m}_{\rm eject}^{(i)} Z_{\rm cold}^{(i)}\,,
\label{eq:hotmetal}
\end{equation}
where again the summation is over all the galaxies in the main halo,
and $\dot{m}_{\rm cool}$ in Eq.(\ref{eq:coldmetal}) is set to be zero
for all but the central galaxies.  We have used the mean 
metallicities, $Z_{\rm cold}=m^{Z}_{\rm cold}/m_{\rm cold}$ 
and $Z_{\rm hot}=m^{Z}_{\rm hot}/m_{\rm hot}$ as the metallicities
for the cold gas in a galaxy, and for the hot gas in a main halo,   
respectively. The ejected metals are assumed to have the same 
fate as the reheated gas.

When solving the above set of differential equations, we divide the time
between two successive snapshots typically into 20 steps. 
As mentioned above, we keep the cooling rate $\dot{m}_{\rm cool}$ 
at a constant level between the two snapshots.

\subsubsection{Halo mergers and galaxy mergers}
\label{sec:merger}

With our high-resolution simulations, we are able to follow the
formation, mass accretion, and mergers of (sub-)halos in detail. 
In our implementation of the SAM, we deal with these processes 
according to the (sub-)halo merger trees at each new snapshot of 
the simulation. For a newly formed halo of mass $m_{\rm vir}$, 
there is hot gas of mass $f_b m_{\rm vir}$ associated with it, 
and a new galaxy is being formed as the gas cools. The accreted 
mass is considered simply by updating the virial mass of the main 
halo (as well as other properties, such as the virial radius, 
circular velocity) and its hot gas in Eq.~(\ref{eq:mhot}). When two 
or more (sub-)halos merge, the hot gas, if any, in the smaller 
halos, is added to that of the main halo that contains them. 
The galaxies in the smaller halos either become halo galaxies 
or satellite galaxies, depending on whether they are the {\it central} 
galaxies of the subhalos.

The merger of a halo galaxy with another halo galaxy or
with the central galaxy is explicitly traced by our subhalo merger trees.  
For a {\it satellite} galaxy where a subhalo is absent in the simulation, 
we assume that it merges with its {\it central} 
or {\it halo}  galaxy after a dynamical friction time scale $\tau$. 
We use the simple formula given in Binney \& Tremaine (1987),
\begin{equation}
\tau = 0.5 \frac {f(\epsilon)V_{c}r_{c}^{2}} 
{C G m_{\rm sat}\ln\Lambda}\,,
\label{eq:dft}
\end{equation}
to calculate $\tau$. As shown by Navarro et al. (1995), this formula
provides a good fit to their numerical simulation studying the
fate of a satellite $m_{\rm sat}$ orbiting in a halo of circular 
velocity $V_{c}$. In the above equation,
$f(\epsilon)$ describes the dependence of $\tau$ on the orbit
eccentricity $\epsilon$, and is approximated by $f(\epsilon) \sim
\epsilon^{0.78}$ (Lacey \& Cole 1993). As in previous SAMs, we adopt
the average value $\langle f(\epsilon)\rangle 
\sim 0.5 $ for $f(\epsilon)$ (e.g. Tormen 1997). 
$C$ in the above equation is a constant approximately equal
to 0.43, and the Coulomb logarithm $\ln \Lambda$
is $\ln (1+m_{\rm vir}/m_{\rm sat})$. We take the satellite 
mass $m_{\rm sat}$ to be the virial mass of the galaxy when it was 
last a central or halo galaxy.

At each new snapshot of the simulations, the merger clock is set up for 
the new {\it satellite} galaxies in each subhalo, according to equation
(\ref{eq:dft}). For those satellites whose host subhalos do not change
between two successive snapshots, the merger clocks are kept
unchanged. In each time step $\Delta t$ in solving the equations
(\ref{eq:mstars})--(\ref{eq:hotmetal}), the merger clock time is 
reduced by $\Delta t$. A {\it satellite} galaxy is 
merged to its {\it central} or {\it halo} galaxy when its merger clock time 
becomes negative, and then is deleted from the galaxy list. 

The physical properties of the merger remnant are modeled according to
the mass ratio of the two galaxies that merge together. Here the mass
of a galaxy is referred to its stellar component only. When the mass
ratio of the smaller galaxy to the bigger one in a merger is less
than 0.3, the merger is called a minor merger; otherwise, it is 
a major merger. In a minor merger, the cold gas of the smaller
galaxy is put to the cold gas disk of the larger one, and its stellar
mass is added to the bulge of the larger one. In a major
merger, all the cold gas in the two galaxies is converted into stars
instantly, and all the stellar mass in the disk and bulge
components is transformed into a new bulge.

\subsubsection{Photometric evolution of galaxies}

We model the photometric evolution of galaxies using the stellar
population synthesis (SPS) model of Bruzual \& Charlot (1993). 
We use their recent SPS library (in 2000) that includes the
metallicity effect on the photometric properties. The spectral energy
distribution of a galaxy can be computed through
\begin{equation}
S_{\nu} = \int^{t}_{0} F_{\nu}(t-t')\dot{m}_{\star}(t')dt'\,,
\end{equation}
where $F_{\nu}$ is the spectral energy distribution for a single-age
population of stars, which depends on the IMF, metallicity
and age of the formed stars. The model of Bruzual \&
Charlot (1993) is used to produce tables for a single
instantaneous burst of a unit stellar mass in which the metallicity
changes from $0.0001$ to $0.1$. We interpolate the tables using a linear
interpolation in $t$ and $\log (Z)$, and calculate the magnitudes 
for both the SDSS filters and the standard Johnson UBV filters.

As in Cole et al. (2000), we introduce a free parameter,
$f_{visible}$, to take into account the fact that part of
the newly formed stars may be invisible 
brown dwarfs ($m \leq 0.1 M_{\odot}$). 
The mass fraction of the visible stars $f_{visible}$ is defined 
through
\begin{equation}
f_{visible} =\frac {\rm mass\, in\, visible\, stars} 
{(\rm mass\, in\, visible\, stars + mass\, in\, brown\, dwarfs)}
\end{equation}
By definition, $f_{visible}$ cannot be larger than $1$. 
The inclusion of brown dwarfs makes the predicted luminosity 
smaller by a factor of $f_{visible}$.  In this paper, we either 
assume $f_{visible}=1$, or when necessary tune the value of $f_{visible}$ 
to match the observed luminosity function in the SDSS $i$ band.
 
\subsubsection{Dust extinction}
\label{sec:dust}
 The dust extinction model we use is similar to that of KCDW. 
We assume that the optical depth in the $B$-band scales with 
luminosity as 
\begin{equation}
\tau_{\rm B}=\tau_{\rm B,\star}
\left(\frac {L_{\rm B}} {L_{\rm B,\star}}\right)^{\beta}
\end{equation}
with $\tau_{\rm B,\star}=0.8$, 
$L_{\rm B,\star}=1.3\times 10^{10}L_{\rm \odot}$ and 
$\beta = 0.5$, as given by Wang \& Heckman (1996). We use the
model of Cardelli et al. (1989) to derive the ratio 
$\tau_{\lambda}/\tau_{\rm B}$. For a thin disk in which 
stars and dust grains have the same distribution, the total 
galactic extinction in magnitude is given by 
Tully \& Fouqu$\acute{e}$ (1985):
\begin{equation}
A_{\lambda}=-2.5 \log_{10} \left (\frac {1-e^{-\tau_{\lambda} \sec \theta}}
{\tau_{\lambda} \sec \theta} \right)\,,
\end{equation}
where $\theta$ is the inclination angle between the disk and the line
of sight. In our model we assume that $\theta$ has a random 
distribution between $0$ and $90$. We apply the above equation 
only for the disk components, and we assume that there is no
dust in the bulge component. As we will see, the inclusion of dust 
extinction improves the luminosity function in the bright end, 
especially for the blue band. Note, however, the dust extinction
adopted here applies only to quiescent disk galaxies, but not 
to dust-enshrouded star burst galaxies. Thus, in our model, all star burst 
galaxies have very blue colors, although in reality they may be rather 
red due to dust extinction. In the present paper we do not 
include dust extinction for starburst galaxies, and we should keep this 
in mind when comparing model predictions with observations. 

\subsection{Model parameters and their impact on model predictions}

Our SAM involves the following set of free parameters:
\begin{itemize}
\item IMF: the initial stellar mass function. In this paper, we
adopt the Salpeter IMF for the major part of our discussion.  
We also consider the Scalo IMF as a comparison;
\item $p$: the yield of metals from a unit mass of newly formed
stars;
\item $R$: the fraction of mass recycled into the cold gas by 
evolved stars;
\item $\alpha_{\rm 0}$: the amplitude of the power-law star formation 
efficiency;
\item $n$: the slope of the power-law star formation efficiency;
\item $\epsilon$: the feedback efficiency;
\item $f_{visible}$: the fraction of newly formed stellar mass that
contributes to the luminosity of a galaxy.
\end{itemize}

 Note that some of the above-listed free parameters are not independent
of each other. For example, for a given IMF, the parameter $R$ is 
determined by stellar evolution theory. For the Scalo IMF, $R$ is about $0.4$, 
and for the Salpeter IMF, $R$ is about 0.35. We will use these values 
for $R$.  The yield $p$ given by stellar evolution theory is 
between 0.01 and 0.02 for both the Scalo and the Salpeter IMFs, 
but the model prediction is still quite uncertain. 
In practice, we constrain $p$ using the color-magnitude relation 
(CMR) of cluster ellipticals, since the CMR is sensitive to 
the changes in the metallicity and the luminosities and 
colors of elliptical galaxies are less affected by dust.

 To fix the other four parameters, we use the following 
two observational inputs: the SDSS galaxy luminosity function in 
the $i$-band, and the cold gas content of the Milky Way.
The luminosity in the $i$-band is neither sensitive to 
the recent star formation rate nor to the dust extinction, and so can be 
used quite reliably as an indicator for the total stellar mass. 
Thus, both the star formation efficiency and the feedback efficiency
are expected to have a significant impact on the $i$-band luminosity 
function and the cold gas content of the Milky Way. 
As in SWTK, we require the predicted cold gas in Milky Way
type galaxies to be about 8$\times 10^{9}\msunhi$.
We will always assume $f_{visible}=1$ (i.e. the mass locked in brown 
dwarfs is negligible), unless a value of $f_{visible}$ less than 1 has 
to be introduced in order to match the observed $i$-band luminosity function 
(only in Fig.\ref{LF_cooling}).

We have adjusted the model parameters by trials and tests. We
find that although the model predictions can be affected 
in complicated ways by changing model parameters, 
the parameter space allowed by the observations is actually quite limited.  
In Table 1, we list the parameters that best match the observational 
inputs. These values of parameters are in good agreement with those 
adopted in many earlier papers on SAMs (e.g. De Lucia et al. 2004).
In what follows we compare model predictions and observational
results based on various properties of the galaxy population.

\section{Comparison with the galaxy population at low redshift}

In this section, we compare our model predictions with observations
about the galaxy population at $z\sim 0$. These observations include
the multi-waveband luminosity functions of galaxies obtained from the
SDSS and the 2dFGRS; the $I$-band Tully-Fisher relation; the CMR of
elliptical galaxies in clusters and in the general field; the
metallicity of galaxies as a function of luminosity or rotation
velocity; and the cold gas fraction in galaxies of different
luminosities; Note that some of the comparisons are not independent
tests, because they are used to calibrate the model parameters. As we
will show, there is general agreement between model and observation
for various properties of the galaxy population.  This is encouraging,
because it indicates that the basic assumptions made in the model may
be not far from reality.  However, we also find significant
discrepancies between model prediction and observation, suggesting
that further improvement has to be made in the SAM.

\subsection {The multi-waveband luminosity functions of galaxies}

\begin{figure}
\plotone{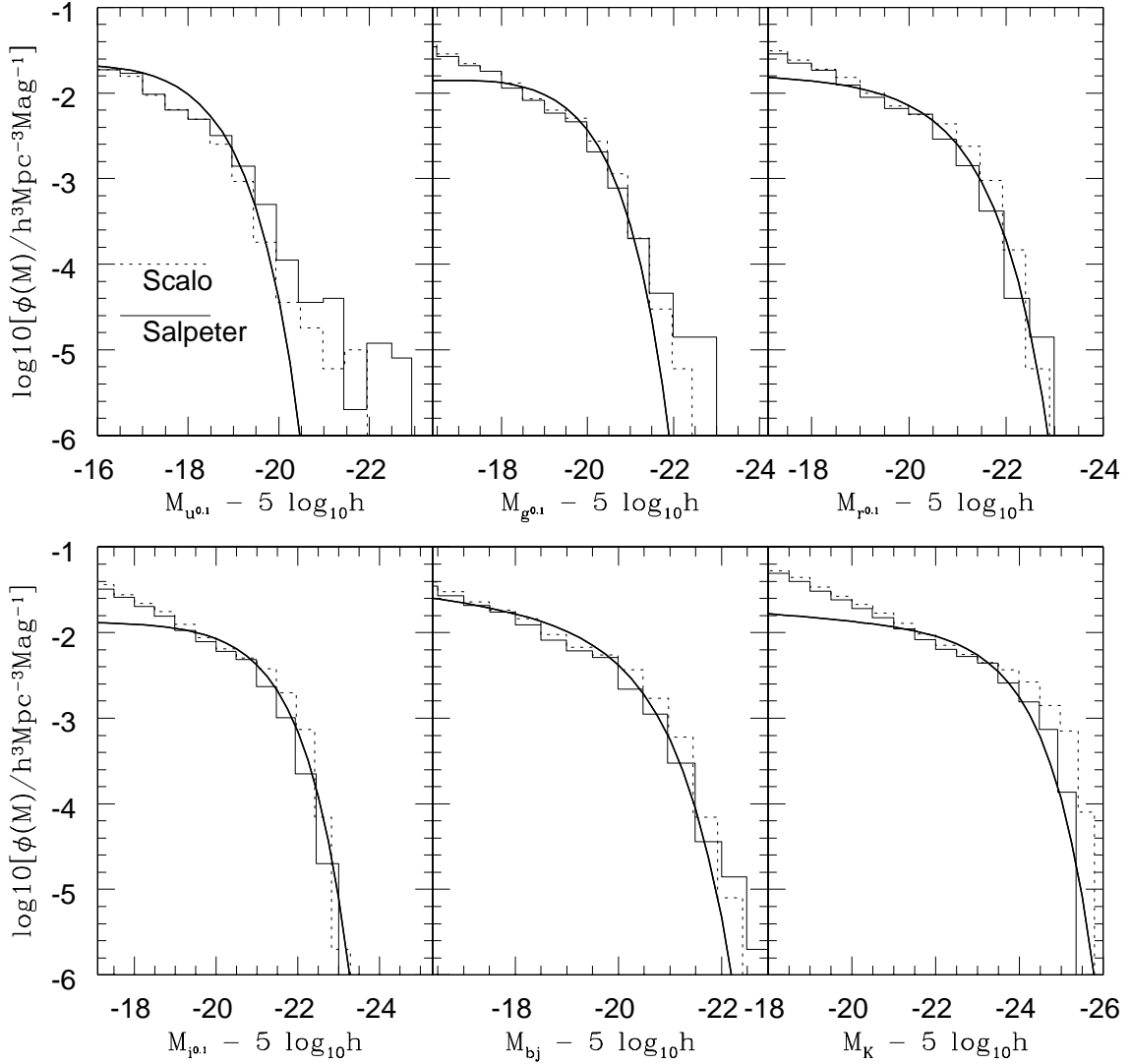} \caption{The multi-waveband luminosity function. The
upper and the lower left panels show the SDSS luminosity functions 
in the  $u^{0.1}$, $g^{0.1}$, $r^{0.1}$, and $i^{0.1}$ bands,
respectively. The lower middle panel shows the result of 
the $b_{j}$-band luminosity function obtained from 2dFGRS, 
and the lower right panel shows the 2MASS $K$-band luminosity function. 
The smooth solid in each panel line show the 
Schechter function fit to the corresponding 
observational result (see text for references). 
In all the panels the histograms show our model predictions 
based on L100 assuming both the Salpeter IMF (solid) 
and the Scalo IMF (dotted).
}
\label{LF}
\end{figure}

Blanton et al. (2003a) estimated the luminosity functions of galaxies
for the SDSS survey in the five SDSS wavebands ($u$, $g$, $r$, $i$ and
$z$). Since the median redshift of the galaxies in their sample is
about 0.1, they made the $K$-correction and evolution correction to
the reference frame at $z=0.1$, and shifted the bandpasses by a factor
of 1.1.  For comparison with these results, we shift the SDSS filters
to the blue by a factor 1.1 when calculating the luminosities of
individual galaxies. Following the SDSS convention, the corresponding
wavebands are denoted by $u^{0.1}$ for the $u$-band, $g^{0.1}$ for the
$g$-band, and so on. The upper panels and the lower left one of
Fig.\ref{LF} present the predicted luminosity functions in the
$u^{0.1},g^{0.1},r^{0.1},i^{0.1}$ bands, with the solid histogram
showing the result for the Salpeter IMF, and the dotted histogram
showing that for the Scalo IMF.  The thick solid lines show the
Schechter function fits to the corresponding SDSS luminosity functions
(Blanton et al. 2003a).  Note that the agreement at the $z^{0.1}$-band
is similar to that of $i^{0.1}$ band, and we have omitted the plot.
We note that these Schechter function fits provide quite accurate
description for the observational data, except at the brightest ends
($M_{\rm r}-5\log_{10} h\approx -22.5$), where they underestimate the
observed number of galaxies.  As one can see from the figure, the
predicted luminosity functions match well the observed ones in the red
bands.  In the lower right panel of Fig.\ref{LF}, we show the
predicted K-band luminosity function.  The smooth solid curve in this
panel shows the result of the Schechter function fit obtained by
Kochanek et al. (2001) from the 2MASS.  This result is similar to that
obtained by Cole et al.  (2001). As we can see, the reasonably good
agreement between model prediction and observation also extends to the
$K$-band, in which galaxy luminosity traces the stellar mass.  Thus,
the SAM considered here can successfully reproduce the observed
stellar mass distribution in the Universe.
 
As a further test, we compare our model prediction with the
observational result of Madgwick et al. (2002) for the 2dF galaxy
luminosity function at the $b_{j}$ band at $z=0$ (see the lower middle
panel of Fig.\ref{LF}).  In our model, we estimate the $b_j$-band
magnitude of a galaxy using the relation $b_{j} = B -0.28(B-V)$ (Blair
\& Gilmore 1982), where $B$ and $V$ are the rest-frame Vega magnitudes
in the $B$- and $V$- wavebands. Since the observed 2dF luminosity
function has a steeper faint-end slope than the SDSS luminosity
function in the blue-band, it is better matched by the model
prediction at the faint end than is the SDSS $g$-band luminosity
function.  From now on, our discussion about the galaxy luminosity
function will be based on results for the $u^{0.1}$, $g^{0.1}$,
$r^{0.1}$ and $i^{0.1}$ bands.

\begin{figure}
\plotone{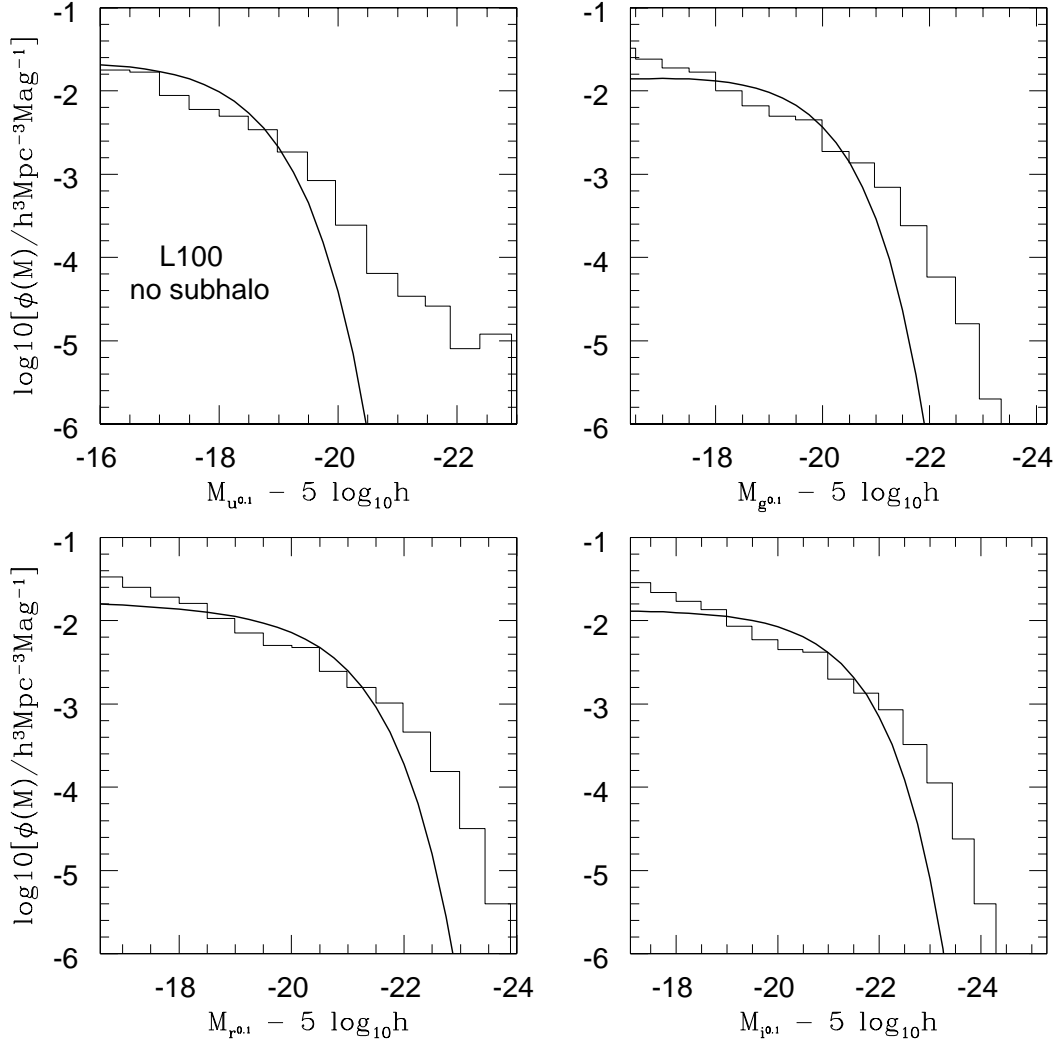} \caption{The same as Fig.(\ref{LF}), 
but here subhalo scheme is not implemented. 
Here we plot the results using the Salpeter IMF.}
\label{LF_Sdss_nosub}
\end{figure}

Compared with previous predictions of the field luminosity functions
based on the SAMs (e.g., KCDW), our model gives a much better match to
the observed luminosity functions.  For example, KCDW significantly
overpredicted the number of galaxies brighter than $L_{\star}$ and
underestimates the number of fainter galaxies.  Consequently, the
luminosity functions predicted in KCDW are much flatter around
$L_{\star}$ than the Schechter function.  A similar discrepancy was
also revealed for the galaxy luminosity function in rich clusters by
SWTK, when a treatment of the galaxy-galaxy mergers similar to KCDW
was adopted.  They attributed the discrepancy to the assumption made
in KCDW on the dynamical friction time scale for satellite galaxies.
With the use of their resolved subhalos in clusters, SWTK found that
the satellite galaxies brighter than $L_{\star}$ in a halo merge into
the central galaxy at a rate much slower than that given by the
dynamical friction time scale of equation (\ref{eq:dft}) (used in
KCDW).  For the same reason, our model can reproduce well the observed
field galaxy luminosity functions at the bright end, because subhalos
hosting bright galaxies are well resolved in our $N$-body
simulations. To demonstrate this, we have considered a model in which
the merger time scales of satellite galaxies are estimated from
equation (\ref{eq:dft}) rather than using the subhalo scheme.  As
shown by Figure \ref{LF_Sdss_nosub}, the luminosity functions
predicted by such a model are much flatter than the observational
results and significantly over-predict the number of bright galaxies.
Although the luminosity function at the bright end can be reduced by
adjusting model parameters, for example by reducing the star formation
efficiency, this will reduce the luminosity function at $L\la
L_{\star}$, making a large discrepancy at the intermediate
luminosity. It is the shape of the luminosity function that cannot be
easily adjusted to match the observations in this model.
 
\begin{figure}
\plotone{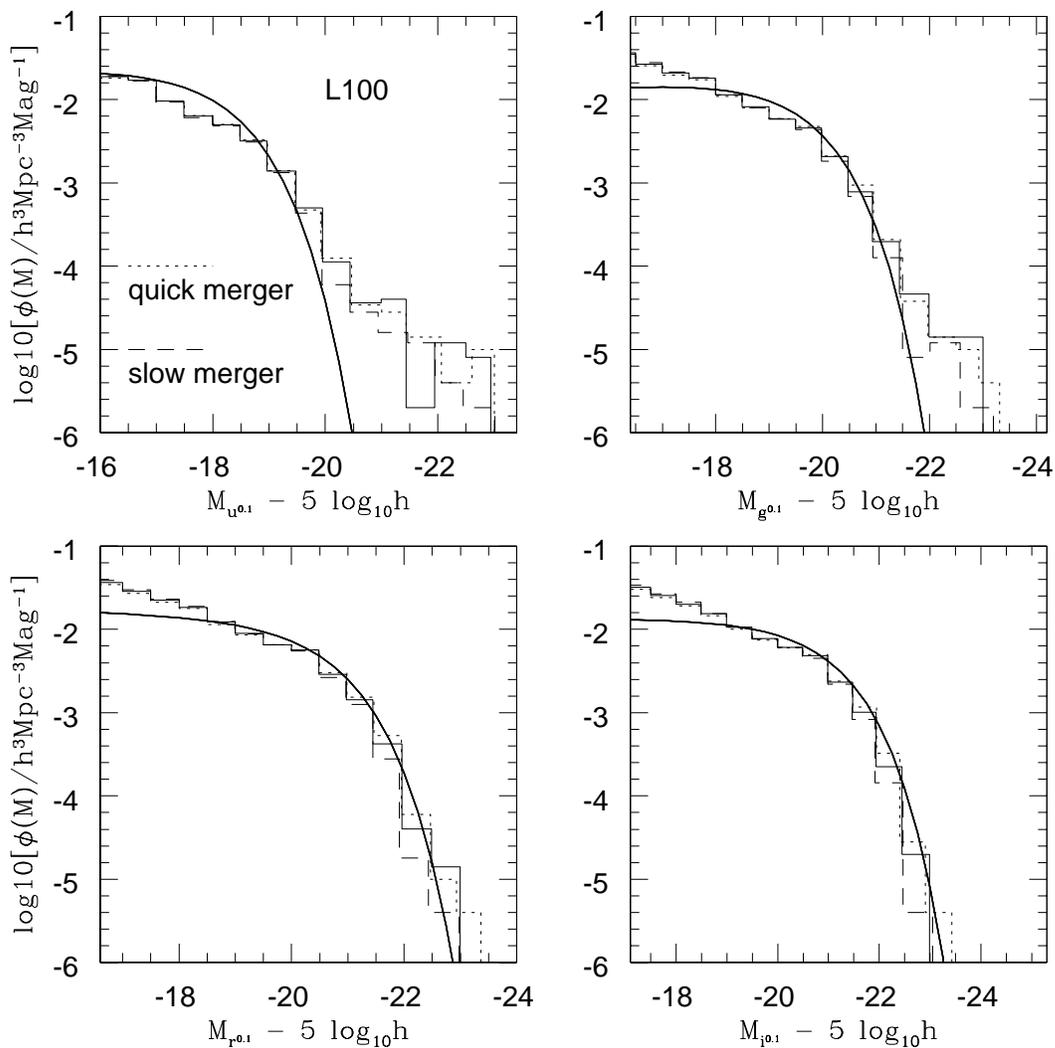} 
\caption{The thick solid lines and the solid histograms 
are the same as in Fig.\ref{LF}.
The dotted histograms result from decreasing 
(\ref{eq:dft}) by half, and the dashed
lines are from increasing the merger time by a factor of 2.}
\label{LF_Sdss_merge}
\end{figure}

 In our subhalo scheme, we have used equation (\ref{eq:dft}) 
to estimate the merger time for the satellites that are currently
not associated with subhalos. It is therefore possible that
the merger time here is also under-estimated. We have tested whether 
the uncertainties of the merger time in such galaxies 
can have a significant impact on our model predictions,
by increasing or decreasing the merger time by a factor 
of 2. The results are shown in Fig.\ref{LF_Sdss_merge}. Little change 
was found in the model luminosity functions. 
The reason is that the merger time of satellites
in these galaxies is relatively long, because of the 
small mass ratio between the satellite galaxy and the central galaxy,
and so mergers are unimportant anyway. 
For the bright end,  the change of the merger timescale 
has only a minor impact, because the mass of a 
satellite galaxy is in general quite small compared
to the bright galaxies.

\begin{figure}
\plotone{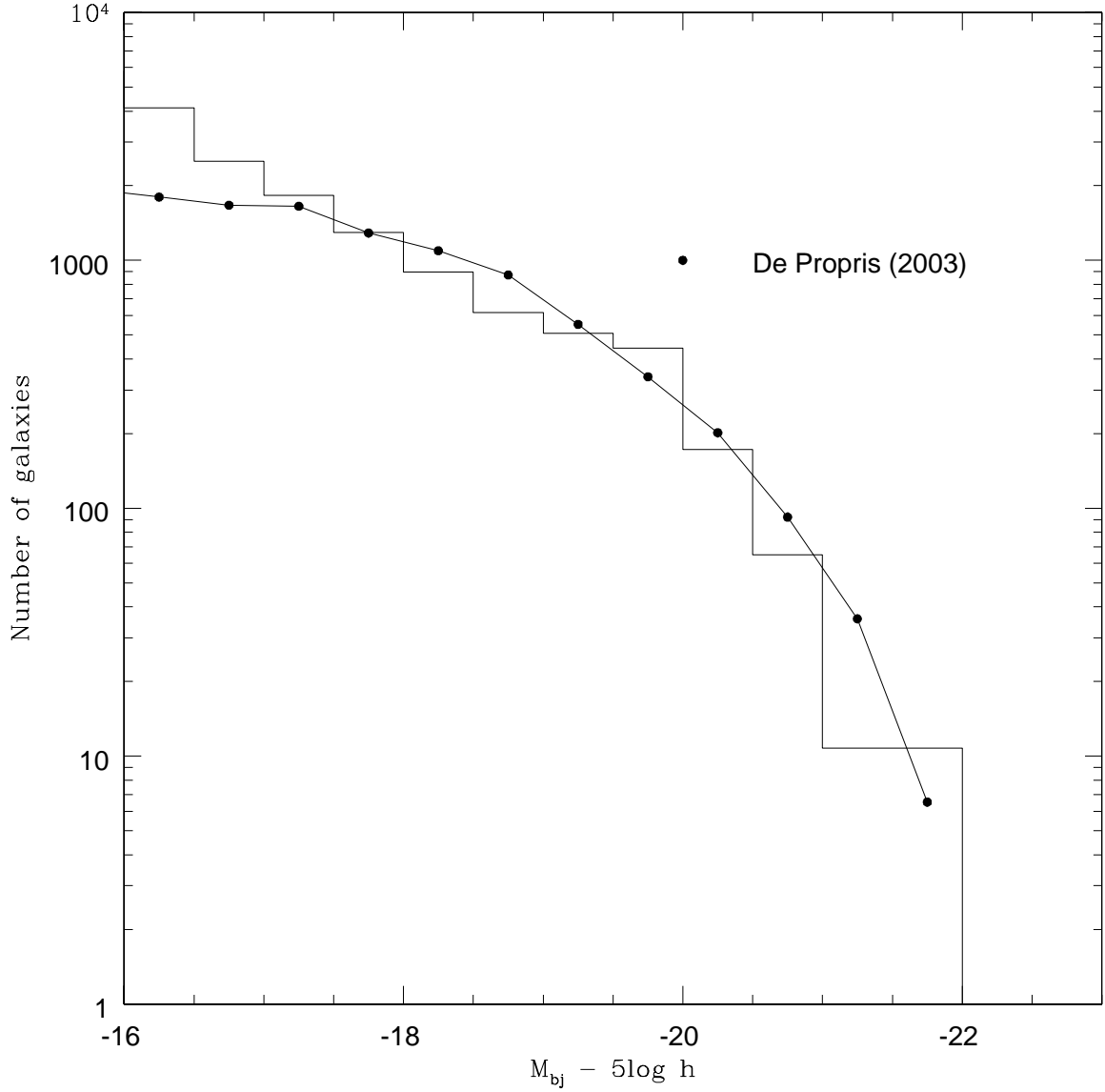}
\caption{The luminosity function of cluster galaxies at the
$b_{j}$-band. The data points connected by a solid line show the
composite luminosity function given by De Propris et al. (2003).}
\label{LF_cluster}
\end{figure}

\begin{figure}
\plotone{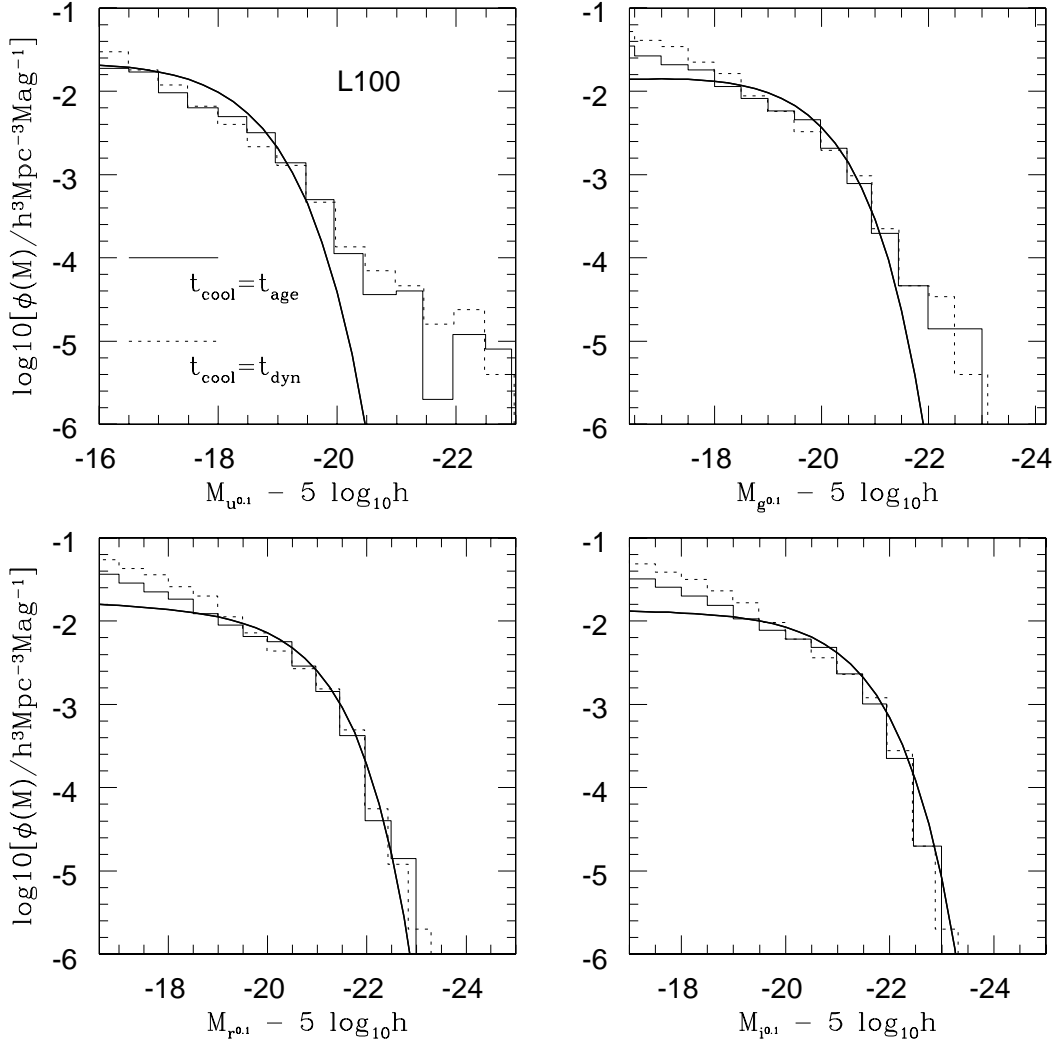} 
\caption{The cooling time effect on the
luminosity function. The solid lines are the same as in Fig1, and
the dotted lines are for $t_{\rm cool} = t_{\rm dyn}$. Note 
that $f_{visible}$ is
$0.5$ for $t_{\rm cool} = t_{\rm dyn}$.}
\label{LF_cooling} 
\end{figure}

In Fig.\,\ref{LF_cluster}, we compare the predicted cluster luminosity
function with the composite luminosity function of rich clusters in
the 2dFGRS by De Propris et al. (2003). Here we use one of our
high-resolution cluster simulations, C1 (other clusters give similar
results). Because the composite LF given by De Propris et al is not
properly normalized for a given cluster mass, we adjust the amplitude
of the observed LF to best match our model cluster LF and compare only
the shape of the LF.  The agreement is quite satisfactory.  The
absolute magnitude of the central galaxy in halo C1 is $M_{b_j} =
-22.0$ in the subhalo scheme, compared with $M_{b_j} =
-24.1$ in the case where galaxy mergers are not based on
subhalos. The subhalo scheme also predicts more sub-$L_{\star}$
galaxies, in better agreement with the observation than the merger
scheme solely based on the dynamical friction time scale. This result
for the cluster LF is very similar to those obtained by SWTK and De
Lucia et al. (2004), both using the subhalo scheme similar to the one
used here. As in the previous works, our result demonstrates that
resolving subhalos is an important ingredient to reproduce the
exponential (Schechter) form at $L_{\star}$ of the composite cluster
luminosity function.

 As emphasized above, the main reason for the success of our 
model in reproducing the luminosity function of field galaxies
is that we are able to follow the trajectories of massive 
galaxies using subhalos in our high-resolution simulations.
This is an important result, because it implies that the 
shape of the observed galaxy luminosity function can be understood 
if galaxy mergers are modeled in a realistic manner.   
There is another factor, which  also contributes to the success 
of our model, namely the cooling time scale we have adopted,
$t_{\rm cool}=t_{\rm age}$ for all halos. 
If we followed SWTK in using $t_{\rm cool}=t_{\rm dyn}$,
we would obtain a much higher amplitude for the luminosity 
function, because the cooling rate for massive halos at 
redshift $0$ is about a factor of $2.5$ higher.
In this case, we have to multiply the luminosity of each galaxy
by $f_{visible}=0.5$,  i.e. to assume that about $50\%$
of the stellar mass is locked into brown dwarfs 
to best match the observed LF 
(see the dotted lines in Fig.\ref{LF_cooling}).
This required $f_{visible}$ is too small to be realistic (Zhao et al. 1996, 
Somerville \& Pirmack 1999).
Also, with the assumption $t_{\rm cool}=t_{\rm dyn}$, the model predicts 
more galaxies at the fainter end, making faint-end  slope of the  
luminosity function of cluster galaxies too steep. 

\begin{figure}
\plotone{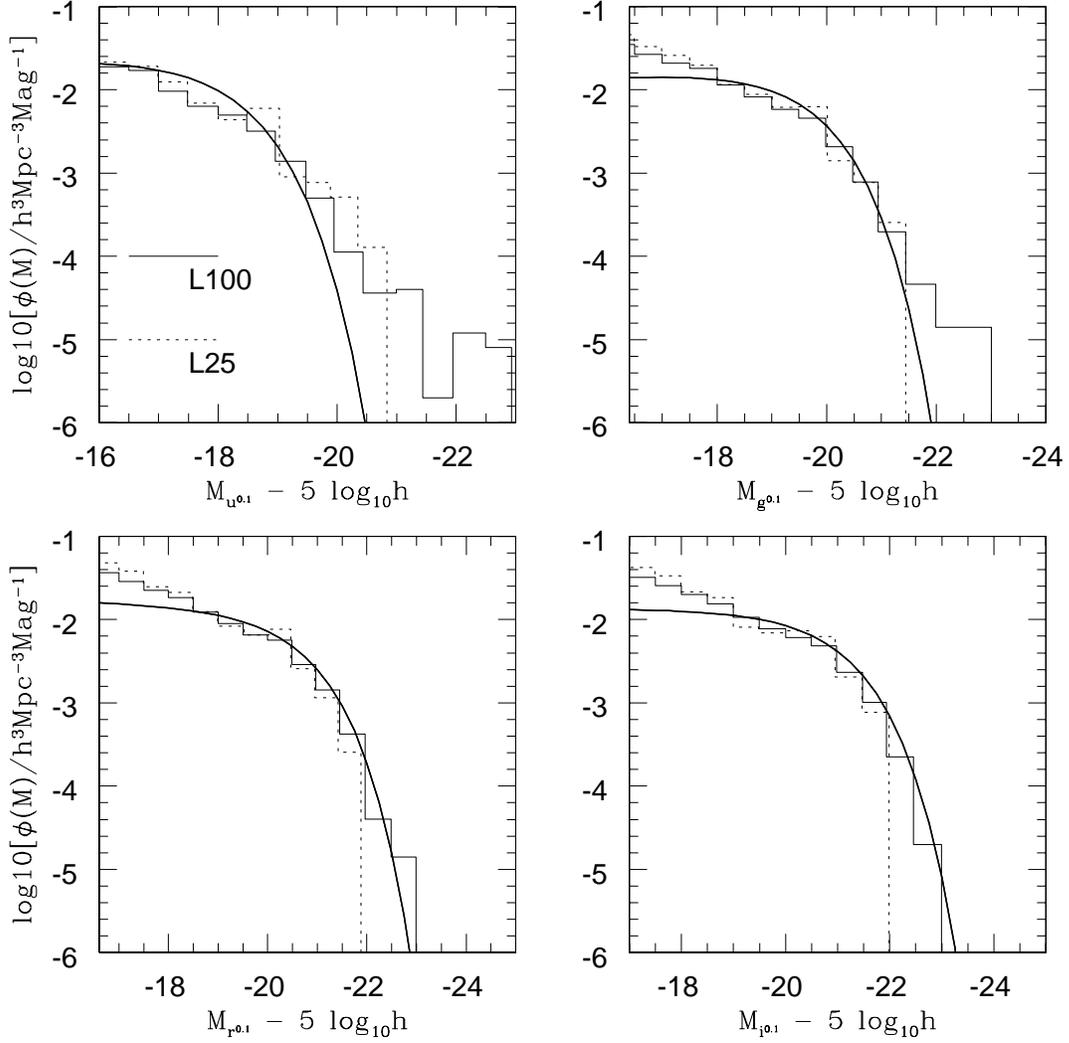} 
\caption{Here we compare the effect of the simulation box on 
the predicted luminosity function for the models L100 and L25.}
\label{LF_Sdss_box}
\end{figure}

As one can see, our model still predicts more faint galaxies
(e.g. $M_g> -18.0 +5\log_{10} h$ in the g-band) than are present in
the SDSS observations. Detailed analysis of these over-predicted faint 
galaxies show that most of them are halo (satellite) galaxies in massive halos. 
Here we want to check whether the model prediction is significantly
affected by numerical resolution.  To do this, we applied the same SAM
to the simulation L25, which has a mass resolution about ten times
higher than L100. Fig.\ref{LF_Sdss_box} shows the results for the L25
simulation (dotted lines) in comparison to the results of L100 (solid
lines).  The resolution improvement does not seem to has any
significant effect (at most $30\%$) on the predicted luminosity function at the faint
end with $M_g > -17.5 +5\log_{10} h$ . As we have shown above, 
the predicted faint end of the luminosity
function is also insensitive to the assumed dynamical friction scale
for satellite galaxies.  The other factor we have examined is the
feedback efficiency.  We have considered a model assuming a feedback
efficiency $\epsilon = 0.2$ instead of $0.1$, and found that such a
strong feedback can bring the number of faint galaxies into agreement
with the observed LF, but the number density of galaxies with $L\sim
L_{\star}$ is reduced too much to be compatible with the
observation.

It is difficult to quantify how serious the faint end problem is.  On
the observational side, the faint-end slope of galaxy luminosity
function is hard to determine, and there are considerable
discrepancies among different sky surveys (LCRS, ESP, SDSS EDR, 2DF,
SDSS DR1, etc. see Driver \& De Propris 2003). Our model prediction is
in good agreement with the steep faint-end slopes observed in the 2dF
survey and in the ESP survey, but it is out of our scope to examine in
which survey the faint galaxies are missed or/and misidentified.

\begin{figure}
\plotone{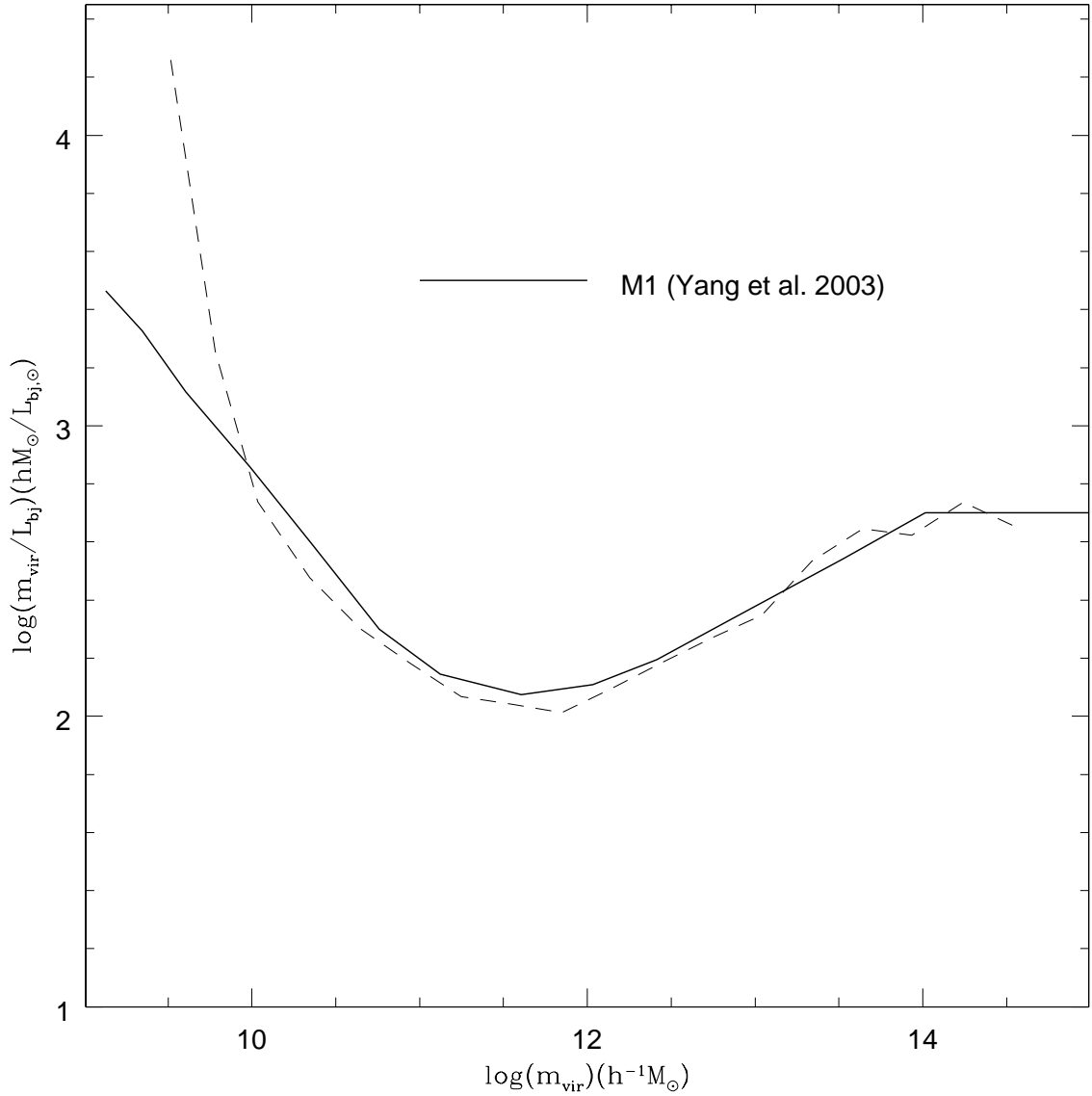} 
\caption{The ratio of the halo mass to the total luminosity $L_{b_J}$
of the halo in our semi-analytical model(the L100 simulation; the
dashed line). For the comparison we also show this function determined
from the 2dFGRS by Yang et al. (2003) using the halo model approach
(the solid line).}
\label{Mass_light}
\end{figure}

Yang et al. (2003) have measured the mass-to-light ratio of dark halos
in the $b_{J}$ band from the 2dFGRS using the halo model
approach. They found that in terms of this ratio, galaxies are formed
most efficiently in the halos of mass $m_{vir}=10^{12}h^{-1}\msun$, and the
galaxy formation is strongly suppressed both in lower and in higher
mass halos. Such suppressions are expected in our semi-analytical
model, because the hot gas cools less efficiently in cluster
halos while the feedback effect can effectively heat up the cold gas 
and hence reduce the star formation in low-mass halos. It is interesting
to see if our model prediction is in quantitative agreement with the
measurement of Yang et al. (2003). This comparison is displayed in
Figure \ref{Mass_light}. Except for $m_{vir}<10^{10}\msunhi$, where the
model prediction is limited by the simulation resolution and the
observational measurement is limited by the small number statistics of
galaxies, the agreement between the model and the observation is
reasonable.

Based on the results obtained above for the $u$-band luminosity
function, we see that our model predicts too many bright blue galaxies
and insufficient number of blue galaxies with luminosities just below
the characteristic luminosity. Our detailed analysis of the model
prediction (with dust extinction) shows that at $z=0.1$ the brightest 
population in the $u$-band (with $M_u$ brighter than $-20.5+5\log h$) 
are dominated by starbursts\footnote{Note that in our model the star 
bursts only occur in major mergers and the recent star-burst galaxies 
have their bursts in the last 0.1G years.} that
have just experienced a major merger. Most of the starbursts are in halos 
with mass around $10^{12}\msunhi$. As we have mentioned before, the
model prediction of the blue luminosities of these galaxies is quite
uncertain, because our model does not include properly dust
extinctions in dust-enshrouded starburst galaxies. The predicted
excess of number of galaxies at lower luminosities ($M_u$ between
$-20.5+5\log h$ and $-19+5\log h$) is mainly due to central galaxies in
dark halos with masses around $10^{13}\msunhi$ (see
Fig.\ref{galaxy_color_mvir}).  In our model, hot gas in such halos is
allowed to cool continuously according to equation (\ref{eq:rcoollh}),
although gas cooling is prohibited in more massive halos (see
\S\ref{ssec:cooling}).  The predicted excess of such blue galaxies is
therefore due to the inadequate treatment of suppressing gas cooling
in massive halos. It is possible to solve this problem by considering
more realistic feedback processes in massive halos.

The cause of the deficit of blue galaxies with luminosities below 
$L_*$ is unclear (also present in Fig.\ref{CM_field}). 
One possibility is that the model assumes
that the star formation rate in a quiescent disk is smooth
with time, while in reality it may be intermittent, consisting
of small bursts of star formation (Kauffmann et al. 2003). 
Such a change does not affect the 
total amount of stars that can form, and so does not change 
the luminosity functions in the red bands, but may produce more 
blue galaxies at the present time, as some of the low-mass
systems may make excursions to the bursting phase at the present time. 
Another possibility is that the IMF is top-heavy, so that the 
fraction of UV-emitting massive stars is increased with respect
to that of the Salpeter IMF. Clearly, further investigations along these 
lines are required to see if they are viable solutions to the problem. 

\subsection{Bimodal distribution in the color-magnitude diagram}

\begin{figure}
\plotone{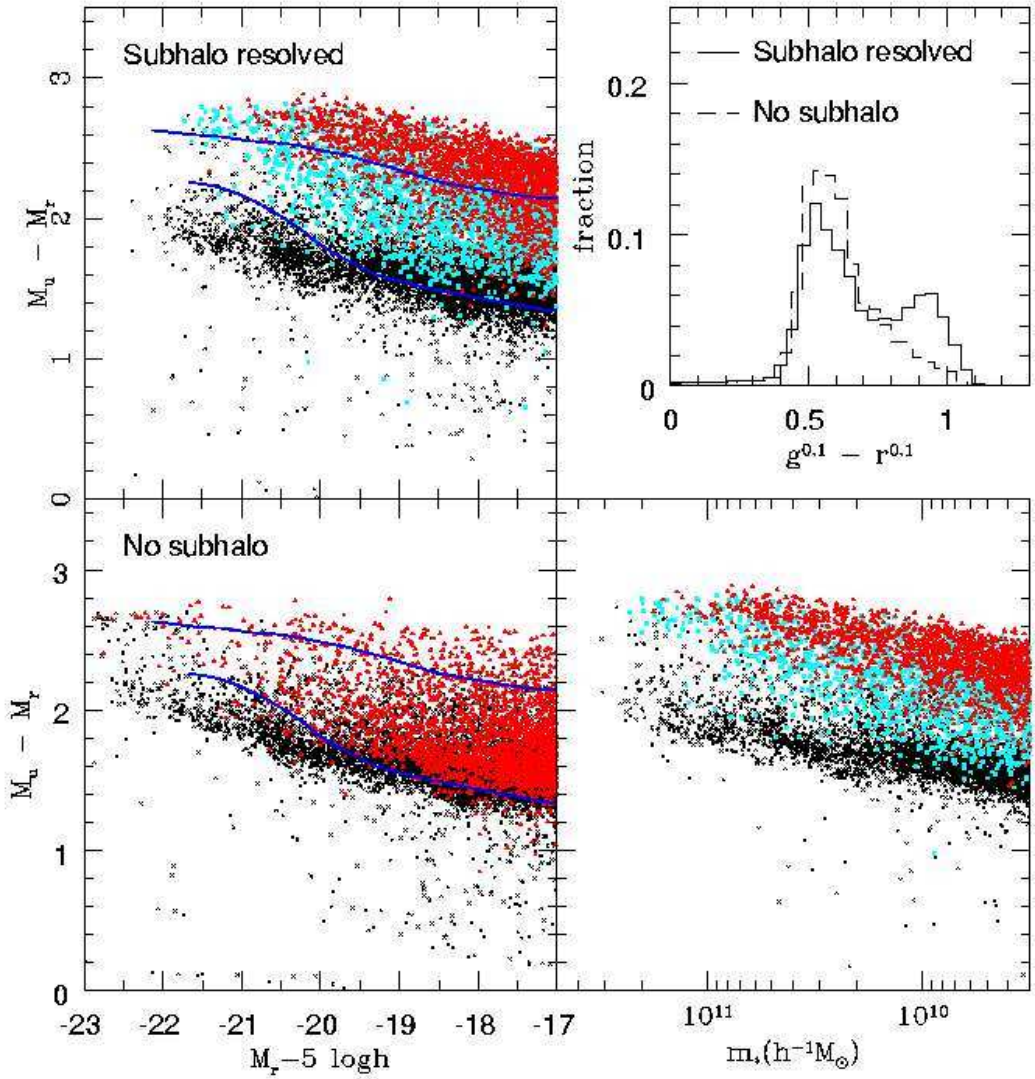} 
\caption{In the left panels: the color-magnitude diagram for all the
galaxies in the L100 simulation at redshift $0$. 
The upper left left panel show results with subhalo resolved
in the simulation and no subhalo resolved in lower panel. The upper right panel
show the $g^{0.1}-r^{0.1}$ color distribution at $z=0.1$ and lower right show the color
vs stellar mass of the model galaxies. In the plot crosses are central 
galaxies, squares are halo galaxies and triangles are satellites.}
\label{CM_field}
\end{figure}

As shown in Baldry et al. (2004a, 2004b), SDSS galaxies seem to 
exhibit a bimodal distribution in the color-magnitude diagram,
in the sense that early-type and late-type galaxies show distinct colors 
for a given magnitude. To see if such a distribution can be reproduced 
in our model, we show in the upper left panel of Fig.\ref{CM_field} the 
$M_u-M_r$ color as a function of $M_{r}$ for our model galaxies 
at redshift $0$. {\it Central, halo} and {\it satellite} galaxies are 
represented by black crosses, cyan squares and red triangles,
respectively. The thick solid lines show respectively 
the fit results to the means of the Gaussian color distributions 
for the red and blue branches obtained by  
Baldry et al. (2004a). For comparison with the results 
of Blanton et al. (2003b), we also show in the upper
right panel the $g^{0.1}-r^{0.1}$ color distribution at $z=0.1$ 
(the solid histogram), which is less dependent on dust extinction
than the $u-r$ color. As in the observations, we see clearly two distinct 
branches in the case where subhalo scheme is used to follow the mergers 
of galaxies, with the blue branch more prominent than the red one. 
In the red branch, most galaxies are early-type galaxies
\footnote{In our model, galaxies are classified into different morphological
types according to the relation between the $B$-band bulge-to-disk 
ratio and the Hubble-type $T$ found by Simien \& de Vaucouleurs (1986).
Following SWTK, we call galaxies with $T< -2.5$ ellipticals, 
those with $-2.5<T<0.92$ S0's, and those with $T>0.92$ spirals and 
irregulars.}. The mean $M_u-M_r$ color in this 
branch is around $2.3$ for faint galaxies ($M_r\sim -17+5\log h$), and 
becomes progressively redder for brighter galaxies. This prediction is 
very close to the observational result, although the predicted 
red branch has a larger dispersion and contains smaller number of 
very bright red galaxies. For the blue branch, 
the predicted mean $u-r$ color for galaxies 
fainter than $-20+5\log h$ is about $1.5$, 
slightly redder than the observational result ($u-r = 1.3$). 
However, the model predicts too many bright galaxies in the 
blue branch. This problem is the same as one see in the $u$-band 
luminosity function shown in Fig.\,\ref{LF}.
The lower right panel shows the $u-r$ color 
versus stellar mass. The behavior seen here is similar to
the color-magnitude relation.

It is instructive to examine what are the main factors that
determine the predicted bimodal color distribution. We have made tests
by changing the prescriptions of gas cooling, dust extinction
and chemical enrichment. These changes can have some effects on the shape 
of the predicted color-magnitude relation, but do not change 
the bimodal behavior. On the other hand, if we use the old
scheme instead of the subhalo scheme to describe the  
galaxy-galaxy mergers, the predicted bimodal feature is 
much weaker (see the lower left panel of Fig.\ref{CM_field}
and the dashed histogram in the upper right panel of the same figure). 
Note that in the upper left panel many of the bright 
galaxies in the red branch are halo galaxies, in which most of 
the cold gas has turned into stars and current star formation 
is at a low level, while in the old scheme no halo galaxies 
exist. The difference between the new and the old schemes can 
be understood as follows. In the subhalo scheme, galaxy mergers 
in a halo are allowed not only with the central galaxy, but also among 
halo and satellite galaxies. Since gas cooling is prohibited in halo and 
satellite galaxies, such mergers involve only evolved stellar population, 
producing new galaxies that are red. Consequently, many galaxies can 
stay red in the subhalo scheme. On the other hand, in the old prescription 
of galaxy-galaxy merging, galaxies in a halo merge relatively fast only 
with the central galaxy that may still have fresh cold gas. 
As a result, the number of red galaxies is reduced and the distinction 
between the red and blue populations is blurred. 
Thus, our ability to follow the evolution of subhalos seems to be
the key in producing the observed bimodal color distribution. 

Baldry et al. (2004a) show in their Figs.3 and 4 that most of the
brightest galaxies are red, that the red branch becomes less prominent
and the blue branch becomes more prominent as luminosity decreases. To
compare our model prediction with their results, we show in
Fig.\ref{Color_strips} the color distributions for galaxies in
different luminosity bins. Overall, the predicted color distributions
are similar to the observed ones, and in most cases the color
distribution resembles the sum of double Gaussian.  However, there is
marked discrepancy between model and observation.  At the
high-luminosity end ($M_{r} - 5 \log h < -19.75$), the predicted
colors in the blue branch are too blue, and the predicted red branch
is too weak.  We believe that this failure is related to our
simplified treatment of the gas cooling in massive halos. As in many
previous works (cf. Kauffmann et al.  1993), we cut off the gas
cooling in very massive halos with the circular velocity above a given
value ($390 \kms$ in our paper).  Although there is strong
observational evidence that the cooling should be suppressed in
massive halos, there is no compelling reason why the gas cooling
should be shut off according to the circular velocity. In fact, the
number of the predicted most luminous red galaxies is very sensitive
to this assumption. If we adopt a toy model in which we shut off the
gas cooling according to the halo mass (instead of the circular
velocity), we find that we can produce more brightest red galaxies and
reduce the number of the brightest blue galaxies, while still keeping the
luminosity function nearly unchanged (Kang et al. 2005). A more
physical model is needed for suppressing the gas cooling in massive
halos.

On the other hand, one also sees that for $M_{r} - 5 \log h >-18.75$,
model galaxies in the blue branch have colors slightly redder than the
observed galaxies.  The observed blue colors imply that these galaxies
have newly formed stellar population (Baldry et al. 2004b; Kauffmann
et al. 2003). This may indicate that the treatment of star formation
in our model is inadequate, and the starbursts triggered by minor
mergers may also be an important missing ingredient for producing more
blue galaxies.

\begin{figure}
\plotone{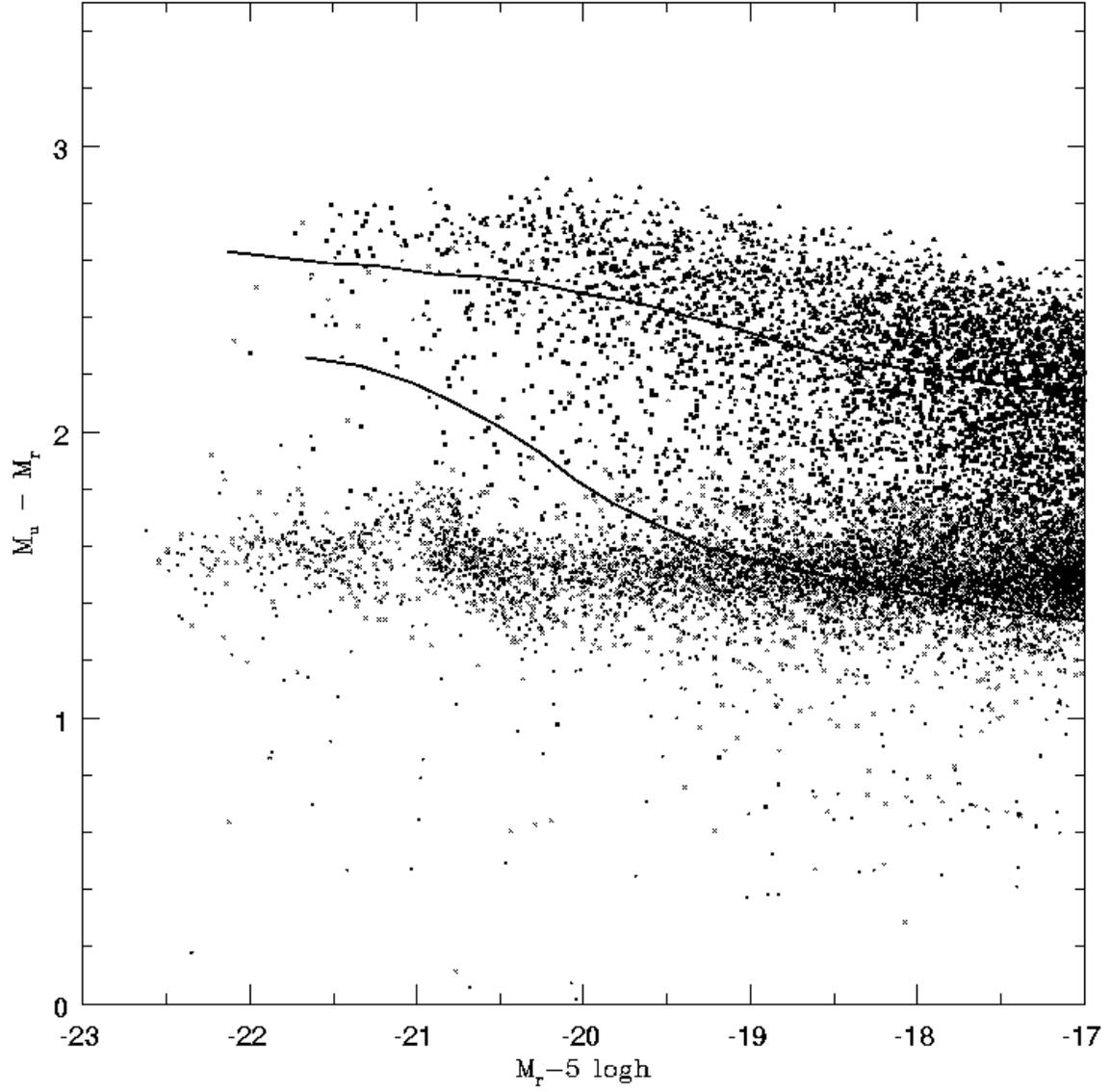} 
\caption{As in upper left panel of Fig.\ref{CM_field} but without colors, and here the dust extinction is not included.}
\label{CM_field_nodust}
\end{figure}

\begin{figure}
\plotone{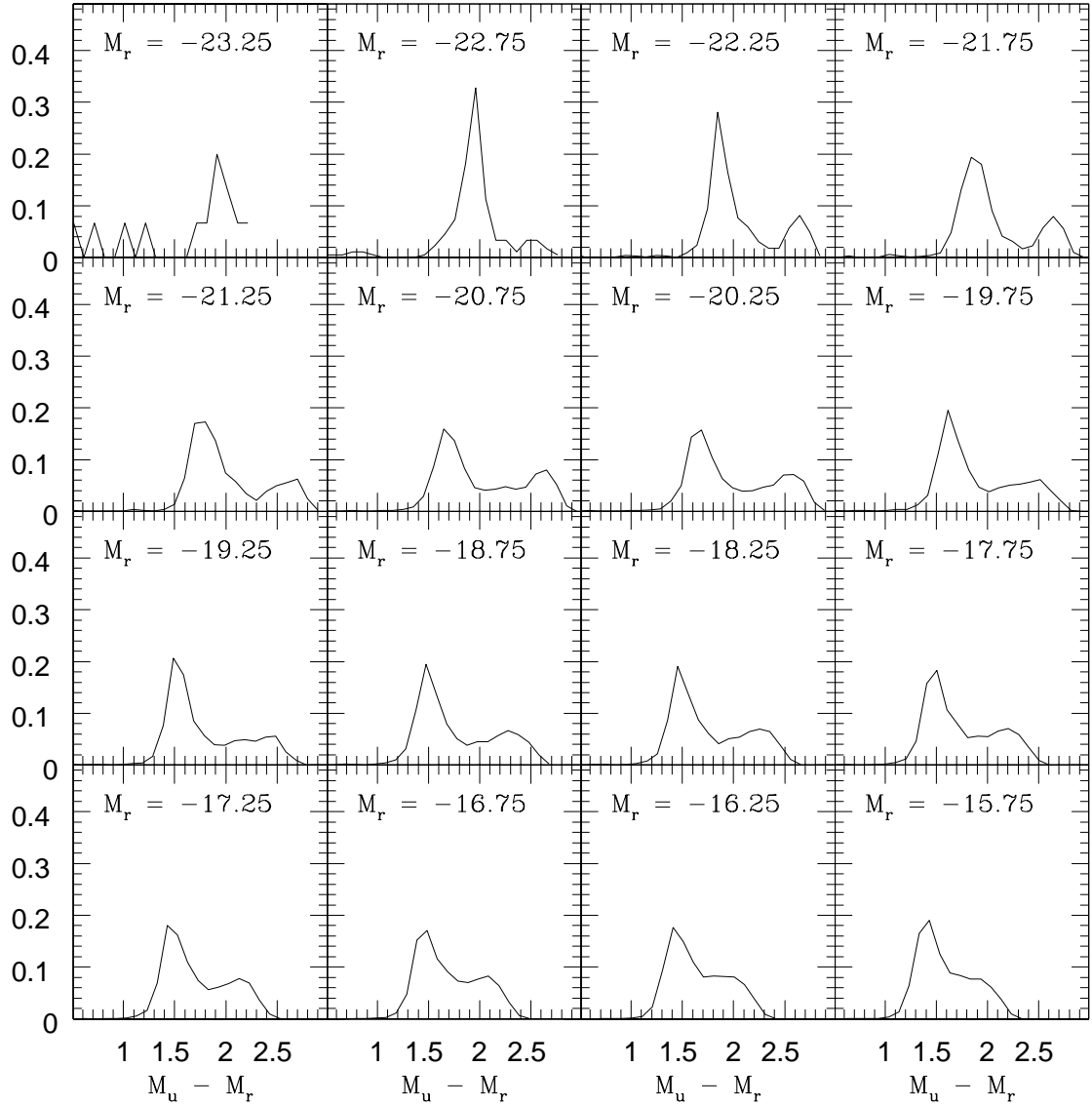}
\caption{The color distributions in absolute magnitude bins of 0.5. In each panel, the label magnitude is the median magnitude.}
\label{Color_strips}
\end{figure}

\begin{figure}
\plotone{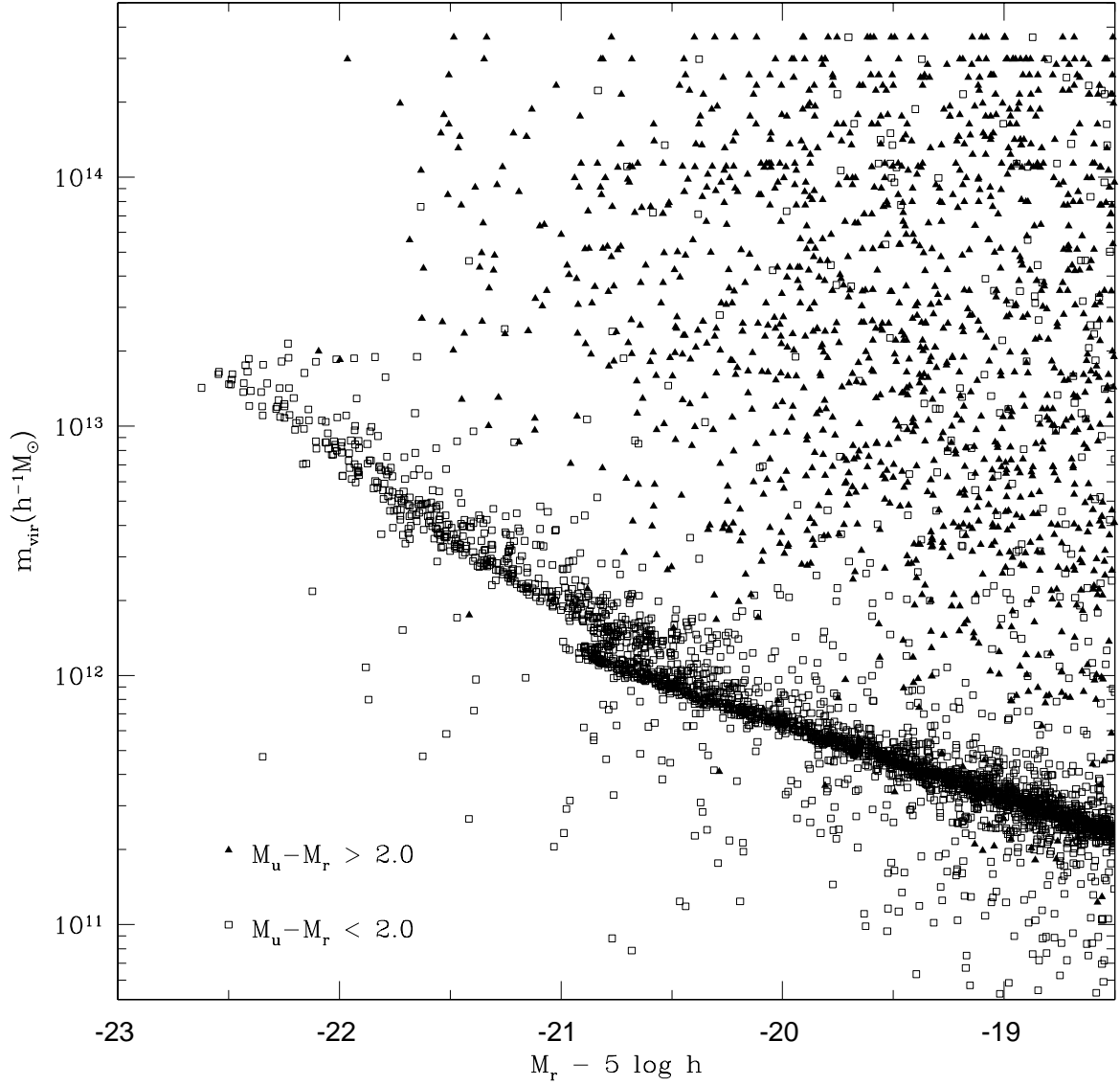} 
\caption{The host halo mass $m_{\rm vir}$ for galaxies of magnitude
$M_{r}$ for blue and red galaxies in the L100 simulation.}
\label{galaxy_color_mvir}
\end{figure}

In our model that includes dust extinction, the predicted colors for galaxies 
with $M_{r} - 5 \log h < -20$ are still bluer than observed. The intrinsic 
colors of these galaxies are very blue, with $M_u - M_r \sim 1.6$, 
as shown in Fig.\ref{CM_field_nodust}. 
In order to see which type of galaxies contribute to the
over-predicted blue galaxies, here we make an census of the galaxy
population in the blue branch.  In our model without dust, about $4\%$
 percent of the very brightest galaxies with $M_{r} \simeq -23.25 $ 
are recent star-burst galaxies in dark halos with virial mass around
$10^{12}\msunhi$. (these are ellipticals because they are associated
with major mergers).  Among the bright galaxies with $M_r<-20 +5 \log
h$, about $10\%$ are E and S0 type galaxies according to the mass
ratio of the merging progenitors and the fraction of star-burst
galaxies is decreased to about $1\%$ percent.  As we have discuss in
\S\ref{sec:dust} that we apply dust extinction correction to the
galaxy with a prominent disk, so the ignore of dust reddening of
star-burst galaxies is not reasonable for many observations show that
dust correction should be important in those systems. The bright blue
galaxies are still over-predicted even though the star-burst galaxies
are extracted.  This is for there are still cold gas available for new
stars forming in the E, S0 and late type galaxies. We also found that
the late type galaxies with bigger disks were also over-predicted
compared with the morphology-classified K-band luminosity function of
2MASS (Kochanek et al. 2001).  In order to see the origin of the
problem, we examine where these blue bright galaxies are located. In
Fig.\ref{galaxy_color_mvir} we therefore plot the host halo mass for
the blue ($M_u-M_r<2.0$) and red ($M_u-M_r>2.0$) galaxies in our L100
simulation. The main population of the blue bright galaxies are the
central galaxies in halos of the velocity dispersion about $350\kms$.
In our model the gas cooling are still allowed in the halos and
therefore the recent newly formed stars make their colors bluer. This
is due to the inadequate treatment of cooling in massive halos, as
discussed in the last subsection. Note that there is a feature in
Fig.\ref{galaxy_color_mvir}, also in Fig.\ref{CM_field_nodust}. This
is caused by the discontinuity of the gas cooling described by equation
(\ref{eq:rcoolsh}) and equation (\ref{eq:rcoollhalo}).

Baldry et al. (2004b) and Balogh et al. (2004) have shown that the 
color bimodality also depends on the environment defined by the local 
density of galaxies. The main properties of their observed 
color distribution can be summarized as follows:
 Firstly, for a given luminosity, the fraction of galaxies 
in the red sequence increases with local density. 
Secondly, faint galaxies are predominantly blue, except 
in very dense environments where a large fraction is also 
red. Finally, the most luminous galaxies are all red, 
independent of local environment. In our model, 
the first two properties are well reproduced. 
However, our model predicts a too strong blue branch 
in low and mediate density environments. 
The main reason for this discrepancy is that 
our model predicts too many luminous blue galaxies, 
because of  the inadequate treatment of gas cooling 
in massive halos and because of the neglect of dust 
attenuation in  massive starburst galaxies.

\begin{figure}
\plotone{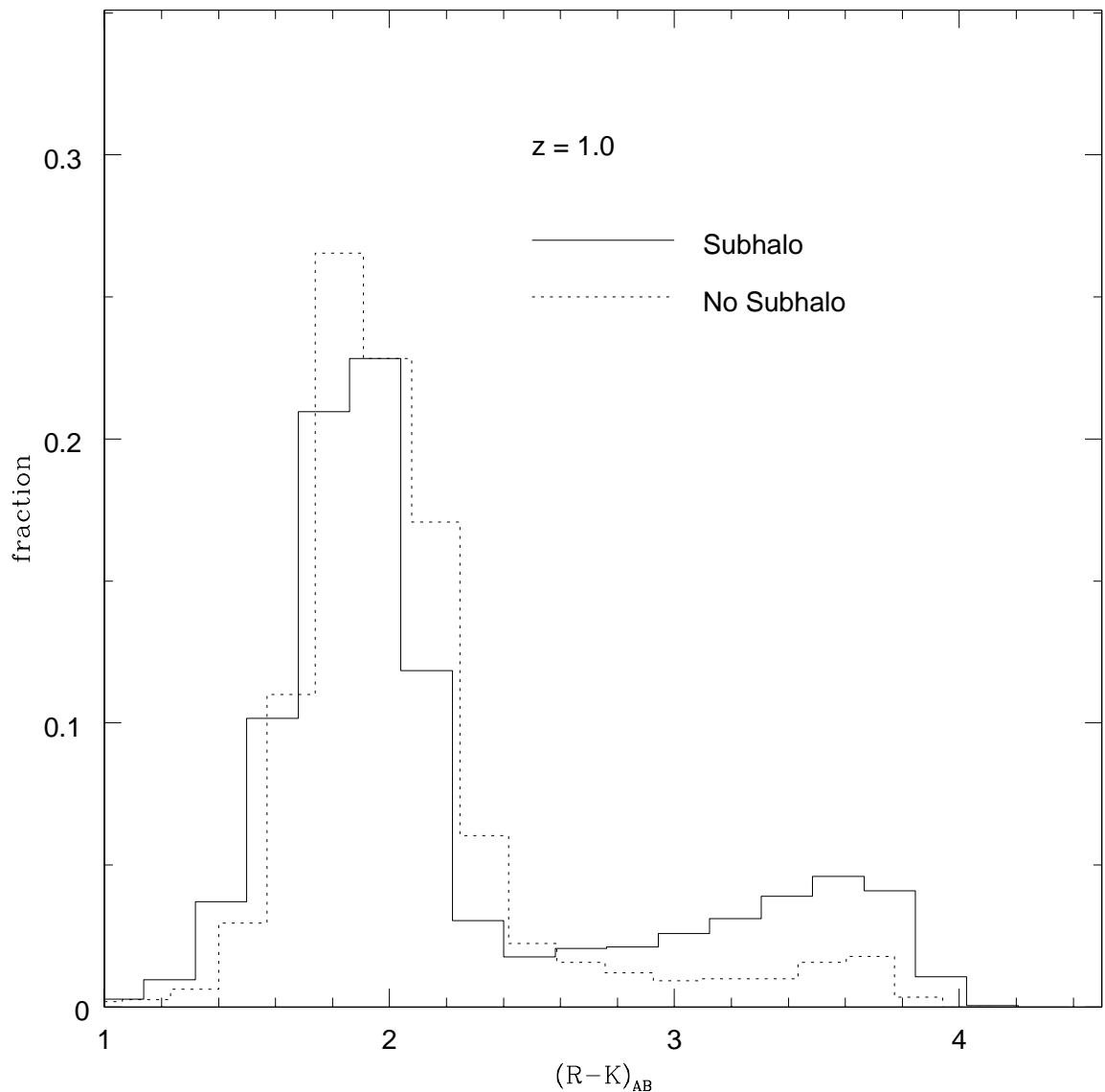} 
\caption{The distribution of $(R-K)$ color for bright galaxies 
with $M_{K} \leq -23.2$ at $z=1$. The $(R-K)$ color 
in the Vega magnitude system was converted into the AB magnitude 
system using $(R-K)_{AB} = (R-K)_{Vega} - 1.65 $.
The solid histogram shows the prediction in which  
subhalo scheme is used to follow the mergers among galaxies, 
while the dotted histogram shows the prediction in which 
mergers of galaxies are based on the dynamical friction 
formula.}
\label{Red_galaxy}
\end{figure}

Since in the subhalo scheme halo- and satellite- 
galaxies can survive longer before merging with the central 
galaxies, our model predicts a large number of bright red galaxies 
at high redshift. In Fig.\ref{Red_galaxy}, we show the distribution 
of galaxies with respect to the observed-frame $R-K$ color at $z=1$. 
Here only bright galaxies with M$_{K} \leq -23.2$ 
were considered; the luminosity limit is chosen to 
match that of the Great Observations Origins Deep Survey (GOODS) sample 
used by Somerville et al. 
(2004) for our adopted cosmology. As shown  
by the solid histogram in Fig.\ref{Red_galaxy}, about 
$10\%$ of our sample galaxies have $(R-K)_{AB}>3.5$. 
If we do not use the subhalo scheme to follow the galaxy-galaxy 
mergers, but use the conventional scheme based on dynamical 
friction, only $3\%$ of the bright galaxies have 
$(R-K)_{AB} > 3.5$ (see the dotted histogram). 

 Based on multi-waveband photometries in deep fields and follow-up 
spectroscopic observations, such as the K20 Survey (Cimatti et al. 2002b), 
the GOODS (Giavalisco et al. 2004), and the Gemini Deep Deep Survey 
(GDDS, Abraham et al. 2004), it is now possible to identify a large 
number of bright galaxies at high redshifts. Such observations 
indicate that the number of bright red galaxies at $z\ga 1$  
may be larger than earlier predictions based on SAM
(e.g. Cimatti et al. 2002a; Somerville et al. 2004; Glazebrook 
et al. 2004). Our results show that resolving subhalos may 
help alleviating this problem. We will make a detailed 
comparison of our model predictions with the observations 
at high redshift in a separate paper. 

\begin{figure}
\plotone{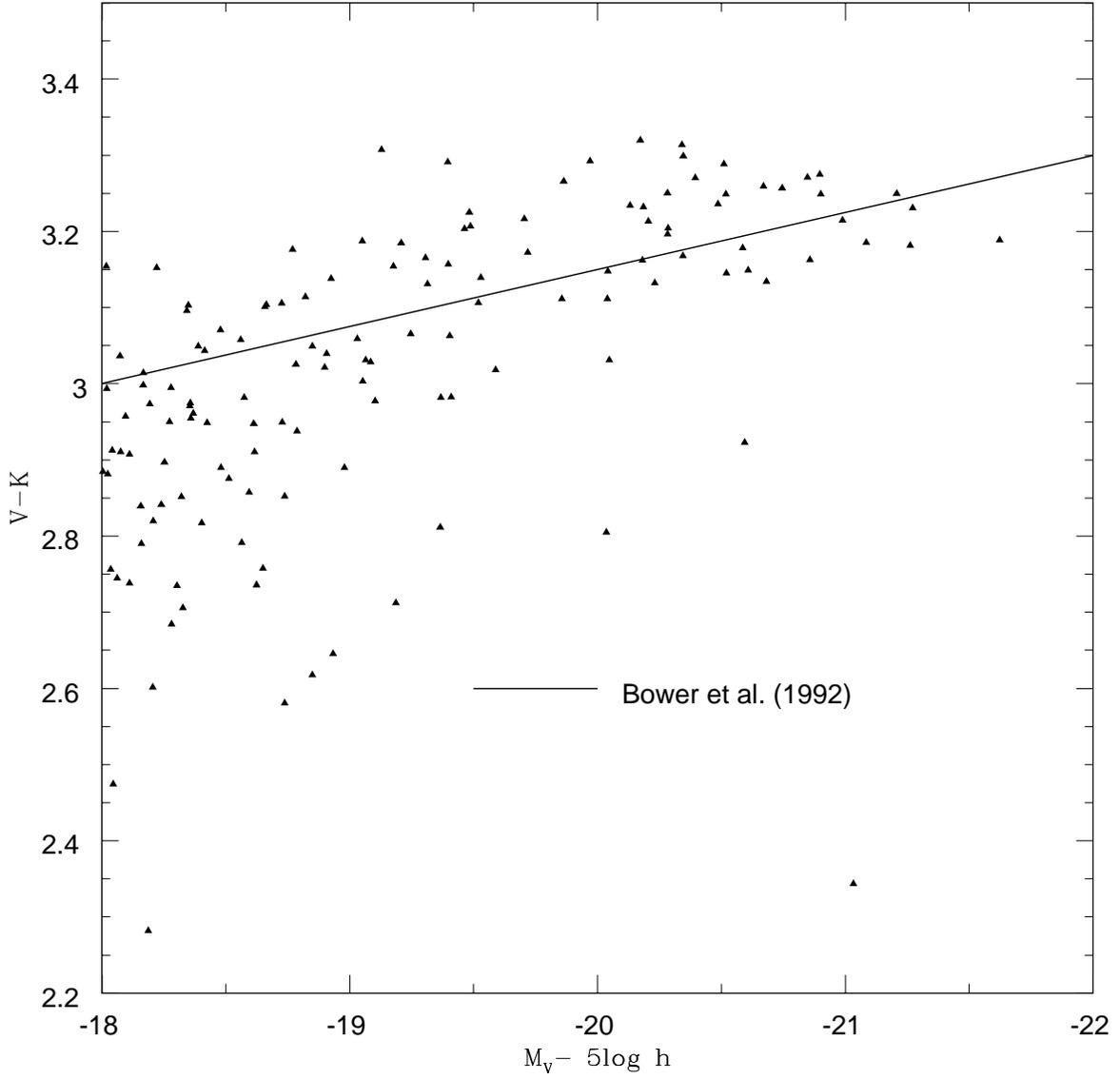}
\caption{Color-magnitude relation for early type galaxies in the C1
cluster simulation. The solid line shows 
the best fit to the observation of Coma
cluster ellipticals by Bower et al. (1992)}
\label{CM_cluster}
\end{figure}

 Before leaving from this subsection, let us consider the
color-magnitude relation for the elliptical galaxies in clusters. In
Fig.\ref{CM_cluster}, we present the predicted color-magnitude
relation for the elliptical galaxies in the cluster simulation C1, and
compare it with observation. We identify elliptical galaxies as the
early-type galaxies with $M_{\rm bulge}-M_{\rm total} <0.8$ in the B
band, and we show the predicted color-magnitude relation for
individual model galaxies as solid triangles. The solid line shows the
fit to the observation of Bower et al. (1992) for elliptical galaxies
in the Coma cluster.  As one can see, the observed trend is well
reproduced in our model.  The main reason for the color-magnitude
relation in the model is the metallicity effect: more luminous
galaxies have redder colors, because they have higher
metallicity. This result is consistent with that obtained by De Lucia
et al. (2004).

\subsection{Tully-Fisher relation}

\begin{figure}
\plotone{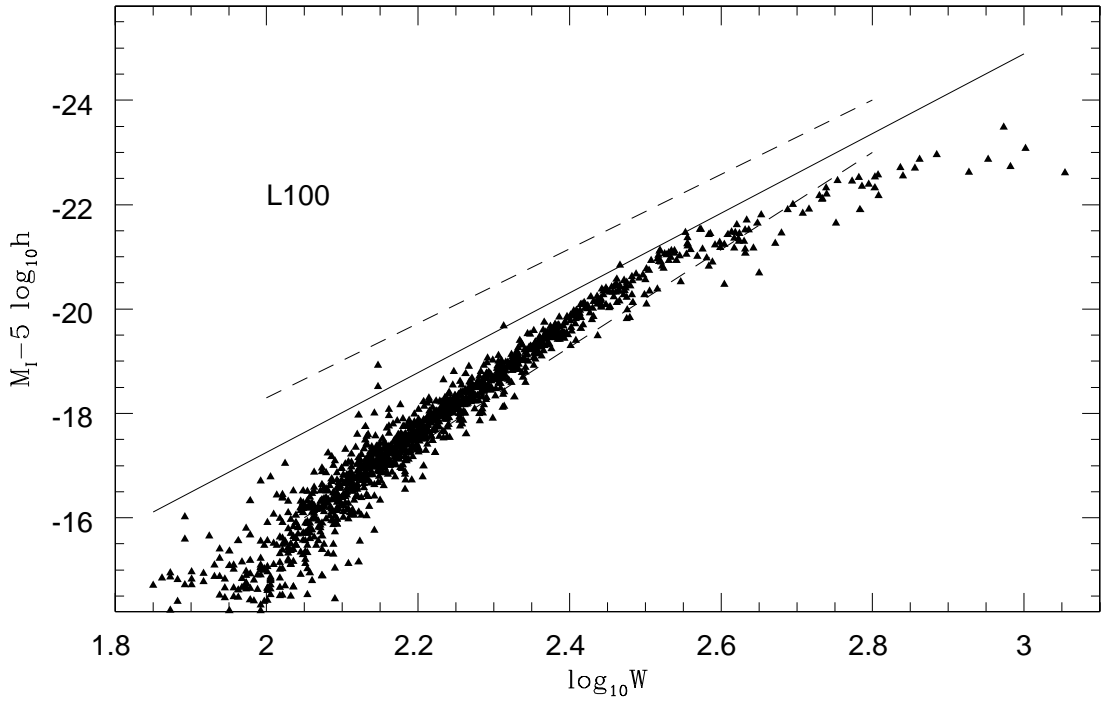}
\caption{The I band Tully-Fisher relation for spiral galaxies in the simulation L100.}
\label{TF}
\end{figure}

In the upper panel of Fig.\ref{TF}, we plot the Tully-Fisher
relation (hereafter TF relation) of the model Sb/Sc field galaxies 
against the observation of Giovanelli et al. (1997) at the I band:
\begin{equation}
M_{I} - 5 \log_{10} h = -21.00 -7.68(\log_{10} W - 2.5)\,.
\label{eq:tf}
\end{equation}
The best fit of the observational result is shown as the solid line,
while the dashed lines show the 1-$\sigma$ scatter. 
To select Sb/Sc galaxies from our SAM, we use the correlation between the
$B$-band bulge-to-disk luminosity ratio and the Hubble-type 
given by Simien \& de Vaucouleurs (1986). We select Sb/Sc 
galaxies according to the criterion, 
$1.2 \leq M_{\rm bulge}-M_{\rm total} \leq 2.5$.
We consider only the central galaxies with such Hubble types
in the {\it main} halos. The velocity width $W$ is set to
be 2 times the maximum circular velocity of the disk, and we have 
assume that the disk maximum circular velocity is $\sim 25\%$ 
larger than the circular velocity of the halo at its virial radius. 
This boost of disk maximum circular velocity  is expected in a 
galaxy halo with typical concentration $c\sim 12$ 
assuming disk mass is negligible (e.g. Mo, Mao \& White 1998). 
The triangles in Fig.\ref{TF} are the results for our model Sb/Sc 
galaxies defined in this way. The figure shows that the 
scatter predicted by the model is significantly smaller 
than in the observational result. Note that in our modeling
we have not taken into account the scatter in the relation 
between the line width and the circular velocity,
neither have we included the observational errors in 
photometry and errors due to dust correction. Both can produce 
scatter in the TF relation. Overall, the predicted 
TF slope agrees quite well with the observed one, but the 
predicted luminosity for a given disk maximum circular velocity
is lower than observed. Note that if dark halo responds to disk 
growth adiabatically, then the boost in the disk maximum circular 
velocity is expected to be larger than what is assumed above,
making the discrepancy between model prediction and observation 
even large. This problem with the Tully-Fisher relation 
in the current $\Lambda$CDM model has been known earlier 
and is due to the fact that galaxy halos predicted by this model are 
too concentrated (e.g. Mo \& Mao 2000 and references therein). 
One possible solution to this problem is that some dynamical 
processes during the formation of galaxies in dark halos
can flatten dark matter halos (e.g. Mo \& Mao 2004), so that 
the boost in the disk maximum circular is reduced. Indeed, if the boost 
is about 10\%, then the predicted Tully-Fisher amplitude 
can match the observation.

\subsection{Metallicity and cold gas fraction in spiral galaxy}
\label{sec:metalcoldgas}

\begin{figure}
\plotone{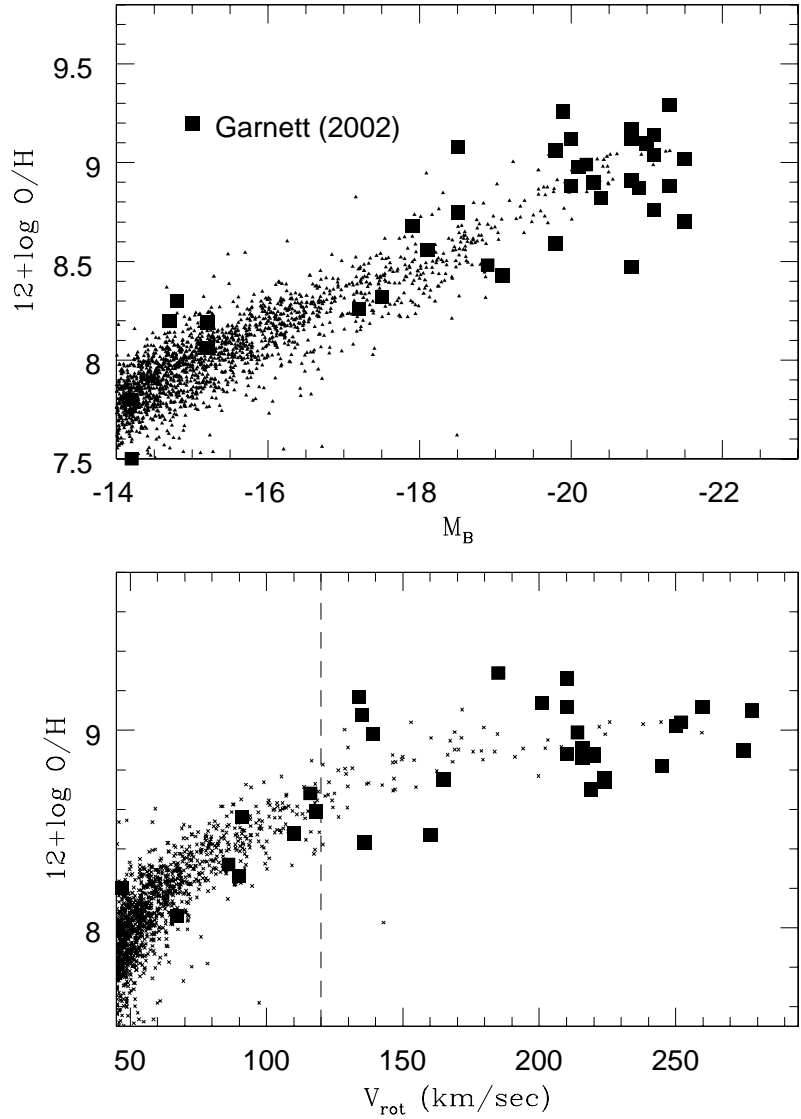} 
\caption{The metallicity of the cold gas as a
function of the luminosity in the B-band (the upper panel) or of the
rotation velocity (the lower panel). The dots are from our SAM, and
the squares are from the observations of Garnett (2002). The dashed line show
the velocity where the metallicity-v$_{rot}$ relation change significantly (Garnett).}
\label{Metal}
\end{figure}

  Garnett (2002) studied the correlation between the metallicity 
of the interstellar gas in a galaxy with the luminosity and rotation
velocity of the galaxy, for a sample of spiral and irregular galaxies. 
In Fig.\ref{Metal}, we compare our model predictions with 
his results. We select a sample of galaxies from 
our SAM with $M_{\rm bulge}-M_{\rm total}>1$, corresponding to 
spiral and irregular galaxies according to our definition.
The metallicity of the interstellar gas is obtained using the 
chemical evolution model described in \S\ref{sec:chemical}.
The figure shows that the observed correlation is well
reproduced by our model. The vertical dashed line in the 
lower panel shows the rotation velocity at which  
the slope of the metallicity-rotation velocity relation 
changes significantly, as pointed out by Garnett.   

\begin{figure}
\plotone{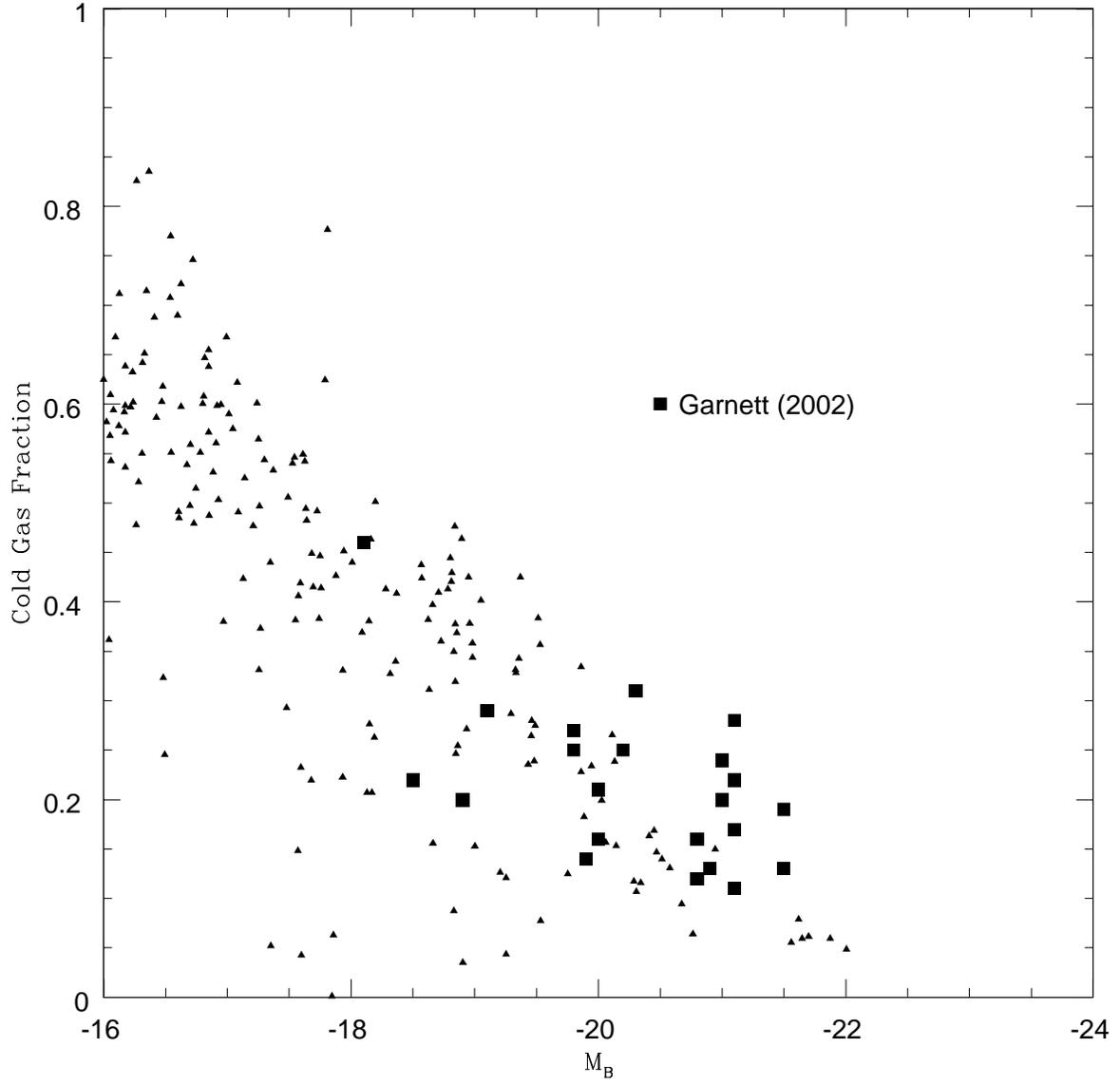} 
\caption{The cold gas fraction as a function of the B-band luminosity. 
The triangles are from our SAM, and the squares are from the 
observations of Garnett (2002).}
\label{ColdGas}
\end{figure}

In Fig.\ref{ColdGas}, we show the cold gas fraction as a function
of galaxy luminosity. Here again we compare our model prediction
with the observational results of Garnett (2002). 
We define the cold gas fraction as $m_{\rm cold}/(m_{\rm cold}+m_{\star})$. 
As one can see, the predicted cold gas fraction is larger 
in fainter galaxies, which is consistent with the observed 
trend, and with the theoretical results obtained by Boissier et al. 
(2001) and  Kauffmann et al. (2003). 
In our model, the predicted cold gas depends sensitively
on the parameter $n$ in equation (\ref{eq:sfr2}).  
Note that $n=0$  gives a constant gas fraction for all the spiral 
galaxies, a result obviously in conflict with the observation. 
A positive $n$ means that smaller galaxies have a lower star 
formation rate, which also leads to the metallicity trends 
shown in Fig.\ref{Metal}.

\section{Discussion and Conclusions}

In this paper, we have modeled galaxy formation in a series of
high-resolution $N$-body simulations of the standard $\Lambda$CDM
model using the semi-analytical approach.  One unique aspect of our
analysis is that our model uses subhalos resolved in the $N$-body
simulations to follow the mergers of galaxies in dark halos. The
semi-analytical model we adopted for galaxy formation in dark matter
halos is similar to those used in recent semi-analytical studies. We
found that, if galaxy-galaxy mergers are followed by the merger
histories of subhalos, rather than based on dynamical friction time
scales, the semi-analytical model can match well the luminosity
functions of galaxies in the bright end in various wavebands redder
than the u-band. Our model also reproduces the main properties of the
observed bimodal color distribution better than that model that do not 
resolve subhalos, although it still predicts too many bright blue
galaxies.  Resolving subhalos also produces many more bright red
galaxies at high $z$, in better agreement with the observations.  Once
the model parameters are calibrated, the model can also match the
observed metallicity-luminosity relation and metallicity-rotation
velocity relation of spiral galaxies, the gas fraction in present-day
spiral galaxies, and the color-magnitude relation for elliptical
galaxies in clusters.

Overall, the model is very successful in explaining a wide variety of
observational facts. This is encouraging, because it means that our
understanding of galaxy formation in the standard $\Lambda$CDM model
is on the right track. However, there are several important issues
that remain unresolved. The first is the process that can effectively
suppress star formation in low-mass halos.  In this paper, instead of
assuming an unrealistically high feedback efficiency, we have
considered a model in which the gas cooling time in a dark halo is
comparable to the age of the system. With such an assumption the
faint-end slope of the luminosity function can be moderated. Although
this assumption leads to predictions that can match the observational
results better, and a prolonged cooling is expected if the
intergalactic medium is preheated (Mo \& Mao 2002, 2004; Oh \& Benson
2003), the details of this process are still unclear. The second is
concerned with the large number of bright blue galaxies predicted by
the model.  To address this issue in detail, we need a realistic dust
model for starburst galaxies, as well as a realistic model for the gas
cooling in massive halos. The third problem is that the current model
predicts insufficient number of blue galaxies with intermediate
luminosities, which indicates that either the IMF in disk galaxies is
top-heavier than the Salpether IMF, or star formation in disk galaxies
is intermittent. Finally, like many earlier analyses based on SAM, we
found that the predicted Tully-Fisher zero point is too low, unless
there are some processes that can significantly flatten CDM
halos. Clearly, further investigations about the possible solutions
are required in order to resolve these issues.

\acknowledgments 

We thank Volker Springel for providing the SUBFIND code, Roberto De
Propris for providing his data in electronic format, Stephane Charlot
and Xu Kong for advice in spectral synthesis model. We also thank
Simon White and Shude Mao for stimulating discussions, and an
anonymous referee for constructive comments. This work is partly
supported in part by NKBRSF(G19990754), by NSFC(Nos.10125314,
10373012, 10203004), and by the CAS-MPG exchange program.  HJM and GB
is supported by the CAS famous scholar program.

\newpage
\clearpage
\begin{deluxetable}{cccccc}
\tablewidth{0pt}
\tablecolumns{6}
\tablecaption{Model parameters in our model}
\tablehead{\colhead{IMF} & \colhead {p} & \colhead {R} & \colhead
{$\alpha_{0}$} & \colhead {$\epsilon$} & \colhead{$n$}}
\startdata
Scalo & 0.04 & 0.4 & 0.10 & 0.20 & 2.2\\
Salpeter & 0.032 & 0.35 & 0.10 & 0.10 & 2.2
\enddata
\end{deluxetable}

\clearpage

\clearpage

\clearpage

\clearpage

\clearpage

\clearpage

\clearpage

\clearpage

\clearpage

\clearpage

\clearpage

\clearpage

\clearpage
\clearpage

\clearpage



\clearpage

\clearpage

\end{document}